\documentclass[a4paper]{llncs}
\usepackage{makeidx}  
\usepackage[linesnumbered,ruled,vlined]{algorithm2e}
\usepackage{amssymb}
\usepackage{amsmath}
\usepackage{graphicx}
\usepackage{multirow}
\usepackage{verbatim}
\usepackage{url}
\usepackage{hyperref}
\usepackage{times}
\usepackage{color}
\usepackage{tikz}
\usetikzlibrary{positioning}
\usepackage{caption}
\usepackage{subcaption}
\usepackage{breakcites}
\usepackage{verbatim}
\begin{document}
\newcommand{\inroundb}[1]{\ensuremath{\left(#1\right)}}
\newcommand{\insquareb}[1]{\ensuremath{\left[#1\right]}}
\newcommand{\inparan}[1]{\ensuremath{\{#1\}}}
\newcommand{\tuple}[1]{\ensuremath{\langle#1\rangle}}
\newtheorem{observation}{Observation}
\newcommand{\Omit}[1]{}
\newcommand{\scomment}[1]{{\color{green}#1}}
\newcommand{\rcomment}[1]{{\color{blue}#1}}
\definecolor{gray}{RGB}{222,222,222}
\newcommand{\highlight}[1]{\colorbox{gray}{#1}}
\pagestyle{plain}
\mainmatter              
\title{Incremental Cardinality Constraints for MaxSAT$^\ddagger$}
%
\titlerunning{Incremental Cardinality Constraints in MaxSAT}  
%
\author{Ruben Martins$^\star$ \and Saurabh Joshi$^\star$ \and Vasco Manquinho$^\dagger$ \and In\^es Lynce$^\dagger$}
\authorrunning{Ruben Martins et al.} 
%
\tocauthor{Ruben Martins, Saurabh Joshi, Vasco Manquinho, Ines Lynce}
\institute{$^\star$University of Oxford, Department of Computer Science, United Kingdom\\
\email{\{ruben.martins,saurabh.joshi\}@cs.ox.ac.uk}\\
$^\dagger$INESC-ID / Instituto Superior T\'ecnico, Universidade de Lisboa, Portugal\\
\email{\{vmm,ines\}@sat.inesc-id.pt}}

\maketitle              

\begin{abstract}
Maximum Satisfiability (MaxSAT) is an optimization variant of the Boolean 
Satisfiability (SAT) problem. 
%
In general, MaxSAT algorithms perform a succession of SAT solver calls to reach an optimum solution making extensive use of cardinality constraints.
Many of these algorithms are non-incremental in nature, i.e. at
each iteration the formula is rebuilt and no knowledge is reused from one iteration to another. 
In this paper, we exploit the knowledge acquired across iterations using novel schemes 
to use cardinality constraints in an incremental fashion. We integrate these schemes 
with several MaxSAT algorithms. Our experimental results show a significant performance 
boost for these algorithms as compared to their non-incremental counterparts.
These results suggest that incremental cardinality constraints could be 
beneficial for other constraint solving domains.

\Omit{
three schemes 
to use cardinality constraints in an incremental fashion: (i) incremental blocking, (ii) incremental 
weakening and (iii) iterative encoding. 
}

\Omit{
Usually, MaxSAT algorithms are based on solving a sequence of closely related 
SAT formulas. Majority of the MaxSAT algorithms are non incremental, i.e. at 
each iteration the formula is rebuilt and no knowledge is reused across iterations. 
%
%
Motivated by the success of incremental SAT solving, 
this paper proposes novel incremental approaches for MaxSAT algorithms. Instead of 
rebuilding the formula at each iteration, we propose to incrementally modify the 
formula while keeping the inner state of the constraint solver.
%
%
In this paper, we show that the proposed incremental approaches 
results in significantly faster MaxSAT algorithms as compared to 
the their non-incremental counterparts. 
%
The success of incremental SAT and MaxSAT solving suggests that
using incremental approaches would be beneficial for any constraint solver.
}

\end{abstract}

\begingroup
   \renewcommand\thefootnote{}
   \footnotetext{$\star$~Supported by the ERC project 280053.\\
   $\dagger$~Partially supported by FCT grants ASPEN (PTDC/EIA-CCO/110921/2009),
POLARIS (PTDC/EIA-CCO/123051/2010), and INESC-ID's multiannual PIDDAC funding
PEst-OE/EEI/LA0021/2013.\\
$\ddagger$The final version of this article is available at \url{http://dx.doi.org/10.1007/978-3-319-10428-7_39}
}
\endgroup

\section{Introduction}
\label{sec:intro}
Plethora of application domains such as 
software package upgrades~\cite{argelich10},
error localization in C code~\cite{jose11}, debugging of hardware
designs~\cite{chen10}, haplotyping with pedigrees~\cite{graca10},
and course time\-tabling~\cite{asin12} 
have benefited from the advancement in MaxSAT solving techniques. 
%
Considering such diversity of application domains for MaxSAT algorithms,
the continuous improvement of MaxSAT solving techniques is imperative.

Incremental approaches have provided a huge leap in the performance of 
SAT solvers~\cite{strichman-charme01,minisat-assumptions,nadel-sat12,audemard-sat13}. 
However, the notion of incrementality has not yet been fully exploited in MaxSAT solving. 
Most MaxSAT algorithms perform a succession of SAT solver calls to reach optimality. 
Incremental approaches allow the constraint solver to retain knowledge from 
previous iterations that may be used in the upcoming iterations.
The goal is to retain the inner state of the constraint solver as well as learned 
clauses that were discovered during the solving process of previous iterations. 
At each iteration, most MaxSAT algorithms~\cite{FM06,manquinho-sat09,msuncore-aaai11,morgado-constraints13} 
create a new instance of the constraint solver 
and rebuild the formula losing most if not all the knowledge that 
could be derived from previous iterations.

Between the iterations of a MaxSAT algorithm, cardinality constraints are added to the formula~\cite{FM06,ansotegui-sat09,msuncore-aaai11,morgado-constraints13}. 
%
Usually, cardinality constraints are encoded in CNF so that a SAT solver can 
handle the resulting formula~\cite{bailleux-cp03,sinz-cp05,asin-constraints11}.
Otherwise, calls to a SAT solver must be replaced with calls to a pseudo-Boolean
solver that natively handles cardinality constraints~\cite{manquinho-sat09}.
This paper discusses the use of cardinality constraints in an incremental manner 
to enhance MaxSAT algorithms.
To achieve this, we propose the following incremental approaches:
(i) incremental blocking, (ii) incremental weakening, and (iii) iterative encoding. 

The remainder of the paper is organized as follows. Section~\ref{sec:preliminaries} introduces preliminaries 
and notations. We describe our proposed techniques in Section~\ref{sec:incremental}. In Section~\ref{sec:related}, 
we mention prior research work done in relevant areas. We show the superiority of our approaches through experimental results 
in Section~\ref{sec:results}. Finally, Section~\ref{sec:conclusions} presents concluding remarks.

\section{Preliminaries}
\label{sec:preliminaries}
A Boolean formula in conjunctive normal form (CNF) is a conjunction of clauses, 
where a clause is a disjunction of literals and a literal is a Boolean variable 
$x_i$ or its negation $\neg x_i$. A Boolean variable may be assigned truth values 
$true$ or $false$. 
A literal $x_i$ ($\neg x_i$) is said to be satisfied if the respective variable is 
assigned value $true$ ($false$). A literal $x_i$ ($\neg x_i$) is said to be unsatisfied 
if the respective variable is assigned value $false$ ($true$). 
A clause is satisfied if and only if at least one of its literals is satisfied.
A clause is called a unit clause if it only contains one literal. 
A formula $\varphi$ is satisfied 
if all of its clauses are satisfied.
The Boolean Satisfiability (SAT) problem can be defined as finding a satisfying 
assignment to a propositional formula $\varphi$ or prove that such an assignment does not exist. 
Throughout this paper, we will refer to $\varphi$ as a set of clauses, where each clause $\omega$ is a set of literals.

Maximum Satisfiability (MaxSAT) is an optimization version of SAT where the goal is to 
find an assignment to the input variables such that the number of 
unsatisfied (satisfied) clauses is minimized (maximized). From now on, 
it is assumed that MaxSAT is defined as a minimization problem. 

MaxSAT has several variants such as partial MaxSAT, weigh\-ted MaxSAT
and weigh\-ted partial MaxSAT~\cite{manya-handbook09}.
A partial MaxSAT formula $\varphi$ has the form $\varphi_h \cup \varphi_s$ 
where $\varphi_h$ and $\varphi_s$ denote the set of hard and soft clauses,  
respectively.
The goal in partial MaxSAT is to find an assignment to the input variables such 
that all hard clauses $\varphi_h$ are satisfied, while minimizing the number of 
unsatisfied soft clauses in $\varphi_s$. The weighted version of MaxSAT allows 
soft clauses to have weights greater than or equal to 1 and the objective is to satisfy 
all hard clauses while minimizing the total weight of unsatisfied soft clauses.
In this paper we assume a partial MaxSAT formula. The described algorithms can be
generalized to the weighted versions of MaxSAT.

Cardinality constraints are a generalization of propositional clauses.
In a cardinality constraint, a sum of $n$ literals must be smaller than or equal
to a given value $k$, i.e. $\sum_{i=1}^n l_i \leq k$ where $l_i$ is a literal. 
As a result, a cardinality constraint over $n$ literals ensures that at 
most $k$ literals can be satisfied.

\Omit{
Although cardinality constraints do not occur in MaxSAT formulations, it is
usual for MaxSAT algorithms to rely on these 
constraints~\cite{FM06,ansotegui-sat09,msuncore-aaai11,morgado-constraints13}.
Usually, cardinality constraints are encoded in CNF so that a SAT solver can 
handle the resulting formula~\cite{bailleux-cp03,sinz-cp05,asin-constraints11}.
Otherwise, calls to a SAT solver must be replaced with calls to a pseudo-Boolean
solver that natively handles cardinality constraints~\cite{manquinho-sat09}.
}


\subsection{MaxSAT Algorithms}
\label{sec:ms_alg}

Due to the recent developments in SAT solving, different algorithms for solving 
MaxSAT have been proposed that rely on multiple calls to a SAT solver.
A SAT solver call \verb;SAT;$(\varphi, \mathcal A)$ receives as input a CNF formula
$\varphi$ and a set of assumptions $\mathcal A$. The set of assumptions 
$\mathcal A$ defines a set of literals that must be satisfied in the model
of $\varphi$ returned by the solver call. Assumptions may lead to early termination
if the SAT solver learns a clause where at least one of the literals in
$\mathcal A$ must be unsatisfied.
An assumption controls the value of a variable for a given SAT call, whereas a
unit clause controls the value of a variable for all the SAT calls after the 
unit clause has been added.
%

The SAT call returns a triple ({\em st}, $\nu$, $\varphi_C$), where {\em st} 
denotes the status of the solver: satisfiable (\textsf{SAT}) or unsatisfiable 
(\textsf{UNSAT}). If the solver returns \textsf{SAT}, then the model that 
satisfies $\varphi$ is stored in $\nu$. On the other hand, if the solver 
returns \textsf{UNSAT}, then $\varphi_C$ contains an unsatisfiable formula
that explains the reason of unsatisfiability. Notice that $\varphi$ may
be satisfiable, but the solver returns \textsf{UNSAT} due to the set of
assumptions $\mathcal A$ (i.e. there are no models of $\varphi$ where
all assumption literals are satisfied). In this case, $\varphi_C$
contains a subset of clauses from $\varphi$ and a subset of assumptions 
from $\mathcal A$. Otherwise, if $\varphi$ is unsatisfiable, then $\varphi_C$ 
is a subformula of $\varphi$.

The algorithms presented in the paper assume that a SAT solver call is
previously performed to check the satisfiability of the set of hard
clauses $\varphi_h$. If $\varphi_h$ is not satisfiable, then the MaxSAT
instance does not have a solution.

\DontPrintSemicolon
\SetKwFunction{soft}{soft}
\SetKwFunction{SAT}{SAT}
\SetKwFunction{decomposeSoft}{partitionSoft}
\SetKwFunction{first}{first}
\SetKwFunction{weight}{weight}
\SetKwFunction{encodeCNF}{CNF}
\SetKwFunction{min}{min}
\SetKwData{result}{satisfiable assignment to}
\SetKwData{unsat}{UNSAT}
\SetKwData{sat}{SAT}
\SetKwData{minc}{min$_\textnormal{c}$}
\SetKwData{true}{true}
\SetKwData{st}{st}
\SetVlineSkip{1pt}
\begin{algorithm}[!t]
  \small
  \KwIn{$\varphi = \varphi_h \cup \varphi_s$}
  \KwOut{satisfying assignment to $\varphi$}
  $(\varphi_W, V_R, \lambda) \gets (\varphi_h, \emptyset, 0)$\;
  \ForEach{$\omega \in \varphi_s$}{
    $V_R \gets V_R \cup \{ r \}$\tcp*[r]{\footnotesize r is a new relaxation variable}
    $\omega_R \gets \omega \cup \{ r\}$\;
    $\varphi_W \gets \varphi_W \cup \{ \omega_R \}$\;
  }
  \While{\true}{
    $(\st, \nu, \varphi_C) \gets \SAT(\varphi_W \cup \{ \encodeCNF( \sum_{r\in V_R} r \leq \lambda ) \}, \emptyset)$ \label{linearus:satcall}\;
    \If{$\st = \sat$}{
      \Return{$\nu$}\tcp*[r]{\footnotesize satisfying assignment to $\varphi$}
    } 
    $\lambda \gets \lambda + 1$\;
  }
  \caption{Linear Search Unsat-Sat Algorithm}\label{alg:linear-us}
\end{algorithm}

Algorithm~\ref{alg:linear-us} performs a linear search on the number of 
unsatisfied soft clauses. First, a new relaxation variable $r$ is added to each 
soft clause $\omega$ (lines 2-5). The goal is to find an assignment to the 
input variables that minimizes the number of relaxation variables 
that are assigned value $true$. 
%
If the original clause $\omega$ is 
unsatisfied, then $r$ is assigned to $true$. At each iteration, a cardinality 
constraint is defined such that at most $\lambda$ relaxation variables 
can be assigned to $true$. This cardinality constraint is encoded into CNF and 
given to the SAT solver (line 7).
Algorithm~\ref{alg:linear-us} starts with $\lambda = 0$ and in each iteration
$\lambda$ is increased until the SAT solver finds a satisfying assignment.
Hence, $\lambda$ defines a lower bound on the number of unsatisfied soft 
clauses of $\varphi$. At each iteration, the result of the SAT call is 
\textsf{UNSAT}, except the last one that provides an optimal solution to 
$\varphi$.

Algorithm~\ref{alg:linear-us} follows an Unsat-Sat linear search.
A converse approach is the Sat-Unsat linear search where $\lambda$ 
is defined as an upper bound. In that case, $\lambda$ is initialized with 
the number of soft clauses. Next, while the SAT call is satisfiable, 
$\lambda$ is decreased. The algorithm ends when the SAT call returns 
\textsf{UNSAT} and the last satisfying assignment found is an optimal 
solution to $\varphi$.

\DontPrintSemicolon
\SetKwFunction{soft}{soft}
\SetKwFunction{SAT}{SAT}
\SetKwFunction{decomposeSoft}{partitionSoft}
\SetKwFunction{first}{first}
\SetKwFunction{weight}{weight}
\SetKwFunction{encodeCNF}{CNF}
\SetKwFunction{min}{min}
\SetKwData{result}{satisfiable assignment to}
\SetKwData{unsat}{UNSAT}
\SetKwData{sat}{SAT}
\SetKwData{minc}{min$_\textnormal{c}$}
\SetKwData{true}{true}
\SetKwData{st}{st}
\SetVlineSkip{1pt}
\begin{algorithm}[!t]
  \small
  \KwIn{$\varphi = \varphi_h \cup \varphi_s$}
  \KwOut{satisfying assignment to $\varphi$}
  $(\varphi_W, \varphi_{W_s}) \gets (\varphi, \varphi_s)$\;
  \While{\true}{
    $(\st, \nu, \varphi_C) \gets \SAT(\varphi_W, \emptyset)$\;
    \If{$\st = \sat$}{
      \Return{$\nu$}\tcp*[r]{\footnotesize satisfying assignment to $\varphi$}
    }
    $V_R \gets \emptyset$\;
    \ForEach{$\omega \in (\varphi_C ~\cap~ \varphi_{W_s})$}{
      $V_R \gets V_R \cup \{ r \}$\tcp*[r]{\footnotesize r is a new relaxation variable}
      $\omega_R \gets \omega \cup \{ r\}$\;
      $\varphi_{W_s} \gets (\varphi_{W_s} \setminus \{ \omega \}) \cup \{ \omega_R\}$\;
      $\varphi_W \gets (\varphi_W \setminus \{ \omega \}) \cup \{ \omega_R \}$\;
    }
    $\varphi_W \gets \varphi_W \cup \{ \encodeCNF( \sum_{r\in V_R} r \leq 1 ) \}$\;
  }
  \caption{Fu-Malik Algorithm}\label{alg:fu-malik-orig}
\end{algorithm}

Core-guided algorithms for MaxSAT take advantage of the certificates of 
unsatisfiability produced by the SAT solver~\cite{FM06}. 
In Algorithm~\ref{alg:fu-malik-orig}, proposed by Fu and Malik~\cite{FM06}, 
soft clauses are only relaxed when they appear in some unsatisfiable 
core $\varphi_C$ returned by the SAT solver. 
Initially, we consider all hard and soft clauses without relaxation. 
In each iteration, an unsatisfiable subformula $\varphi_C$ is identified and 
relaxed by adding a new relaxation variable to each soft 
clause in $\varphi_C$ (lines 7-11). Additionally, a new constraint is added 
such that at most one of the new relaxation variables can be assigned 
to $true$ (line 12). The algorithm stops when the formula becomes satisfiable.

\DontPrintSemicolon
\SetKwFunction{soft}{soft}
\SetKwFunction{SAT}{SAT}
\SetKwFunction{decomposeSoft}{partitionSoft}
\SetKwFunction{first}{first}
\SetKwFunction{weight}{weight}
\SetKwFunction{encodeCNF}{CNF}
\SetKwFunction{min}{min}
\SetKwData{result}{satisfiable assignment to}
\SetKwData{unsat}{UNSAT}
\SetKwData{sat}{SAT}
\SetKwData{minc}{min$_\textnormal{c}$}
\SetKwData{true}{true}
\SetKwData{st}{st}
\SetVlineSkip{1pt}
\begin{algorithm}[!t]
  \small
  \KwIn{$\varphi = \varphi_h \cup \varphi_s$}
  \KwOut{satisfying assignment to $\varphi$}
  $(\varphi_W, V_R, \lambda) \gets (\varphi, \emptyset, 0)$\;
  \While{\true}{
    $(\st, \nu, \varphi_C) \gets \SAT(\varphi_W \cup \{ \encodeCNF( \sum_{r\in V_R} r \leq \lambda ) \}, \emptyset)$ \label{msu3:satcall}\;
    \If{$\st = \sat$}{
      \Return{$\nu$}\tcp*[r]{\footnotesize satisfying assignment to $\varphi$}
    }
    \ForEach{$\omega \in (\varphi_C ~\cap~ \varphi_s)$}{
       $V_R \gets V_R \cup \{ r \}$\tcp*[r]{\footnotesize r is a new variable}
       $\omega_R \gets \omega \cup \{ r\}$\tcp*[r]{\footnotesize $\omega$ was not previously relaxed}
       $\varphi_W \gets (\varphi_W \setminus \{ \omega \}) \cup \{ \omega_R \}$\;
    }
    $\lambda \gets \lambda + 1$\;
  }
  \caption{MSU3 Algorithm}\label{alg:msu3}
\end{algorithm}

In Algorithm~\ref{alg:fu-malik-orig} soft clauses may have to be relaxed 
several times. As a result, several relaxation variables can be added to 
the same soft clause. Nevertheless, other 
core-guided algorithms have already been proposed where at most one relaxation
variable is added to each soft clause~\cite{ansotegui-sat09,wmsu3-corr07}.
Algorithm~\ref{alg:msu3} follows a linear search Unsat-Sat, but
soft clauses are only relaxed when they appear in some unsatisfiable 
core $\varphi_C$.

In this section we solely describe MaxSAT algorithms that will be the 
focus of the enhancements proposed in the paper. We refer to the literature
for other approaches such as branch and bound algorithms using MaxSAT
inference techniques or procedures to estimate the number of unsatisfied 
clauses to prune the search~\cite{manya-handbook09}.
Additionally, there is also an extended overview on core-guided 
algorithms~\cite{morgado-constraints13}.

\subsection{Totalizer Encoding}
\label{sec:cardinality}
\Omit{
Let $x_1,\dots,x_n$ be a set of Boolean variables of interest. A cardinality  constraint
over these variables of the form $x_1 + \dots + x_n \leq k$, where $1\leq k\leq n$
indicates a constraint that at most $k$ of these $n$ variables be set to $true$. Encoding
such a constraint as a CNF formula in an efficient manner is not a trivial task and many
different encodings have been proposed (CITE) with varying degree of efficiency.
}

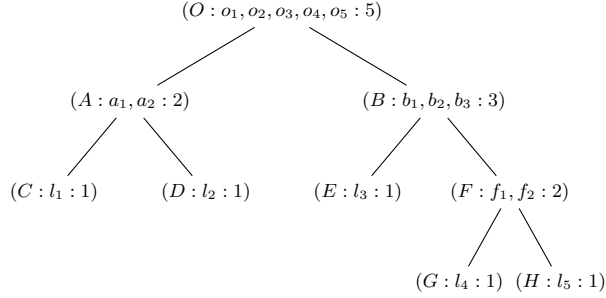
\begin{figure}[!t]
\centering
\begin{tikzpicture}[level/.style={sibling distance=50mm/#1},scale=0.8, every node/.style={scale=0.8}]
\Omit{\node (a) {$(A:o_1,o_2,o_3,o_4,o_5:5)$}
	child { node (b) {$(B:s^1 _1,s^1 _2:2)$ } 
		child { node (d) {$(D:l_1:1)$} } 
		child { node (e) {$(E:l_2:1)$} } }
	child { node (c) {$(C:s^2 _1, s^2 _2, s^2 _3:3)$} 
                child { node (g) { $(G: l_3:1)$}}
		child { node (f) {$(F:s^3 _1, s^3 _2:2)$} 
			child { node (h) {$(H:l_4:1)$}}
			child { node (i) {$(I:l_5:1)$}}} };}
\node (a) {$(O:o_1,o_2,o_3,o_4,o_5:5)$}
	child { node (b) {$(A:a_1,a_2:2)$ } 
		child { node (d) {$(C:l_1:1)$} } 
		child { node (e) {$(D:l_2:1)$} } }
	child { node (c) {$(B:b_1, b_2, b_3:3)$} 
                child { node (g) { $(E: l_3:1)$}}
		child { node (f) {$(F:f_1, f_2:2)$} 
			child { node (h) {$(G:l_4:1)$}}
			child { node (i) {$(H:l_5:1)$}}} };
\end{tikzpicture}

\caption{Totalizer encoding for $l_1 + \dots + l_5 \leq k$}
\label{fig:totalizer}
\end{figure}

For the purpose of this paper, we describe the Totalizer encoding~\cite{bailleux-cp03} 
for cardinality constraints, as later in the paper we build upon this encoding to 
present our novel approaches. 
Totalizer encoding can be better visualized as a tree as shown in
Fig.~\ref{fig:totalizer}. Here, notation for every node is
$(node\_name:node\_vars:node\_sum)$. To enforce the cardinality constraint, we need to
count how many input literals $(l_1,\dots,l_n)$ are set to $true$. This counting is done
in unary. Therefore, at every node its corresponding $node\_vars$ represents an integer from
$1$ to $node\_sum$ in the order. For example, at node $B$, $b_2$ being set to $true$
means that at least two of the leaves under the tree rooted at $B$ have been set to $true$.
The input literals $(l_1,\dots,l_5)$ are at the leaves where as the root node has the output variables $(o_1,\dots,o_5)$
giving the finally tally of how many input literals have been set.

Any intermediate node $P$, counting up to $n_1$, has two children $Q$ and $R$
counting up to $n_2$ and $n_3$ respectively such that $n_2+n_3=n_1$. Also, their
corresponding $node\_vars$ will be $(p_1,\dots,p_{n_1})$, $(q_1,\dots,q_{n_2})$ and
$(r_1,\dots,r_{n_3})$ in that order. In order to ensure that the correct sum is 
received at $P$, the following formula is built for $P$:

\begin{equation}
\bigwedge _{
\begin{tiny}
\begin{array}{c}
0 \leq \alpha \leq n_2 \\ 0 \leq \beta \leq n_3 \\ 0 \leq \sigma \leq n_1 \\ \alpha +
\beta = \sigma
\end{array}
\end{tiny}
} \neg{q_\alpha} \vee
\neg{r_\beta} \vee p_\sigma \quad \text{where,}\ p_0=q_0=r_0=1
\label{eq:totalizersum}
\end{equation}

Essentially, Eq.~\ref{eq:totalizersum} dictates that if $\alpha$ many leaves have been set
to $true$ under the subtree rooted at $Q$ and $\beta$ many leaves have been set to $true$
under the subtree rooted at $R$ then $r_\sigma$ must be set to $true$ to indicate that at
least $\alpha + \beta$ many leaves have been set to $true$ under $P$.
Eq.~\ref{eq:totalizersum} only counts the number of input literals set to
$true$. In other words, it encodes \emph{cardinality sum} over input literals. To enforce that at most $k$ of the input literals are set to $true$, we conjunct it
with the following :
\begin{equation}
 \bigwedge _{k+1 \leq i \leq n} \neg{o_i}   
\label{eq:comparator}
\end{equation}


\begin{observation}
Two disjoint subtrees for the Totalizer encoding are independent of each other.
For example, the tree rooted at $B$ counts how many literals have been set from
$(l_3,l_4,l_5)$ where as, the tree rooted at $A$ counts the set literals from $(l_1,l_2)$.
\label{obs:independent}
\end{observation}

Note also that Eq.~\ref{eq:totalizersum} counts up to $n$ and then
Eq.~\ref{eq:comparator} restricts the sum to $k$. If we only want to enforce the
constraint for at most $k$ then we need at most $k+1$ output variables at the root.
In turn, we need at most $k+1$ $node\_vars$ at any intermediate node. Even with this
modification, Eq.~\ref{eq:totalizersum} remains valid. However, the equality $n_2+n_3=n_1$ may no
longer hold. With this modification,  Eq.~\ref{eq:comparator}
simplifies to \[ \neg{o_{k+1}} \]
Without the simplification this encoding requires $O(nlog\ n)$ extra variables and
$O(n^2)$ clauses.  After the simplification the number of clauses reduces to 
$O(nk)$~\cite{buttner-icaps05,qmaxsat-jsat12}. From here on, we will refer to this 
simplification as \emph{$k$-simplification}.

\begin{observation}

Let $\varphi_1$ and $\varphi_2$ be two formulas, representing cardinality sums $k_1$ and $k_2$ respectively, generated using Eq.~\ref{eq:totalizersum} and $k$-simplification. Observe that $\varphi_1 \subset \varphi_2$, whenever $k_1 < k_2$.

\Omit{
Let $\varphi_1 = \textnormal{CNF}_(l_1 + \ldots + l_n \leq k_1)$ 
and $\varphi_2 = \textnormal{CNF}(l_1 + \ldots + l_n \leq k_2)$ be two cardinality 
constraint formulas encoded into CNF with the Totalizer encoding.
We say that $\mathbb{E}$ is a monotonic cardinality encoding iff $\varphi_1 \subset \varphi_2$, whenever $k_1 < k_2$.
}
\label{obs:monotonic}
\end{observation}

\Omit{
If we only want to enforce the
constraint for at most $k$ then we need at most $k+1$ output literals at the root.
In turn, we need at most $k+1$ $node\_vars$ at any intermediate node. Even with this
modification, the Eq.~\ref{eq:totalizersum} remains valid. However, the equality $n_2+n_3=n_1$ may no
longer hold.
}

\Omit{
With this modification,  Eq.~\ref{eq:comparator}
simplifies to:\[ \neg{a_{k+1}} \]
}

\Omit{
Eq.~\ref{eq:totalizersum} can be simplified as follows : 
\begin{equation}
\bigwedge _{
\begin{tiny}
\begin{array}{c}
0 \leq \alpha \leq min(n_2,k+1) \\ 0 \leq \beta \leq min(n_3,k+1) \\ 0 \leq \sigma \leq
min(n_1,k+1) \\ \alpha +
\beta = \sigma
\end{array}
\end{tiny}
} \bar{q_\alpha} \vee
\bar{r_\beta} \vee p_\sigma \quad \text{where,}\ p_0=q_0=r_0=1
\label{eq:totalizersumk}
\end{equation}
}

\section{Incremental Approaches}
\label{sec:incremental}
\Omit{
To enhance MaxSAT algorithms that are based on refining unsatisfiable SAT 
formulas, we propose the following incremental approaches based on assumptions: 
(i) blocking, (ii) incremental weakening, and (iii) iterative encoding. 
}

MaxSAT algorithms that are based on refining unsatisfiable SAT formulas can be 
enhanced by changing cardinality constraints in an incremental fashion. In this
section, we propose the following three techniques to enable incrementality when
using cardinality constraints: 
(i) incremental blocking, (ii) incremental weakening, and (iii) iterative encoding. 


\subsection{Incremental Blocking} 
MaxSAT algorithms based on refining unsatisfiable formulas are usually 
non-incremen\-tal.
After an unsatisfiable iteration, the formula is refined by 
removing a certain set of clauses and adding a new set of clauses that imposes a 
weaker constraint over the relaxation variables. 
However, SAT solvers do not allow the deletion of clauses that belong 
to the original formula. Since learned clauses from previous iterations may depend on the
clauses that are now being removed, it is not sound to keep all of the
learned clauses. Incremental SAT solving addresses these problems by using 
assumptions~\cite{minisat-assumptions}. To the best of our knowledge this approach has not 
been extended for incremental MaxSAT solving.

\Omit{
Incremental blocking is based on extending the use of assumptions to 
disable some of the clauses from previous MaxSAT iterations and to enhance MaxSAT algorithms 
with incrementality.}

We denote $b$ as a \emph{blocking variable} which is used to extend a clause $\omega$ to $(\omega \vee b)$.
When $b$ is set to $false$ the original clause $\omega$ is enforced (enabled). When $b$ is set to $true$
the extended clause $(\omega \vee b)$ is trivially satisfied and $\omega$ is no longer enforced (disabled).
Thus, adding $b$ (or $\neg b$) as an assumption or unit clause 
disables (or enables) a clause. Using a blocking variable, we can overcome the limitation of a SAT solver not allowing 
clause deletions.

\Omit{
\begin{definition}[Incremental Blocking]
Let $\varphi$ be a MaxSAT formula.
When adding a clause $\omega$ to $\varphi$ that may be removed in the 
future, we extend $\omega$ with a fresh variable $b$, i.e. $(\omega \vee
b)$. 
If the clause is enabled, then we add $\neg b$ to the set of assumption variables. 
If the clause is disabled, then we add $\ b\ $ as a unit clause to $\varphi$.
From here on, we will refer to this fresh variable $b$ as a blocking variable. 
\end{definition}
}

\Omit{
\scomment{I think assumptions and unit clauses should be defined in preliminaries}
Assumptions control the value of a given variable in the current iteration, whereas 
unit clauses control the value of a given variable in all upcoming iterations. 
In MaxSAT algorithms if a clause is disabled at a given iteration, 
then it will not be enabled in any upcoming iteration. Note that a disabled 
clause is always satisfied due to assignment of the blocking variable.
}

\subsubsection{MaxSAT Algorithms based on Cardinality Constraints.}
Many MaxSAT algorithms are based on refining the formula by encoding and updating  
cardinality constraints~\cite{msuncore-aaai11,ansotegui-cp13,morgado-constraints13}. 
For these algorithms, the incremental blocking can be done when cardinality 
constraints are encoded to CNF.

\begin{subequations}
\begin{align}
\varphi \boxplus b \equiv \inparan{\omega \vee b : \omega \in \varphi} \label{eq:extend}\\
\Psi(\mathbf{X},k,b) \equiv \mathtt{CNF}_{Tot^k}(\Sigma x_i)\boxplus b  \label{eq:cardextend} \\
\varphi^i \equiv \varphi_W \cup \inroundb{\bigcup^{i} _{j=1} \Psi(\mathbf{X^j},k^j,b^j)} \cup \tuple{\neg b^i,\neg o_{k^i+1}} \cup \insquareb{b^1,\dots,b^{i-1}} \label{eq:incblock-i}\\
\varphi^{i+1} \equiv \varphi_W \cup \inroundb{\bigcup^{i+1} _{j=1} \Psi(\mathbf{X^j},k^j,b^j)}\cup \tuple{\neg b^{i+1},\neg o_{k^{i+1}+1}} \cup \insquareb{b^1,\dots,b^i} \label{eq:incblock-inext}
\end{align}
\end{subequations}

Let Eq.~\ref{eq:extend} define the extension of a CNF formula $\varphi$ with a 
blocking variable $b$. Next, $\Psi(\mathbf{X},k,b)$ represents a cardinality sum 
up to $k+1$ over $x_1,\dots,x_n$ encoded in CNF using Eq.~\ref{eq:totalizersum} 
and $k$-simplification of the Totalizer encoding and extended with a blocking variable $b$. 
Then, for incremental blocking, at line~\ref{linearus:satcall} in 
Algorithm~\ref{alg:linear-us} and line~\ref{msu3:satcall} in Algorithm~\ref{alg:msu3} 
we call the solver on $\varphi^i$ as defined in Eq.~\ref{eq:incblock-i} for 
the $i^{th}$ iteration. Assumption $\tuple{\neg b^i}$ enables the cardinality 
constraint for the current iteration whereas unit clauses $\insquareb{b^1,\dots,b^{i-1}}$ 
ensure that cardinality constraints from earlier iterations are disabled. 
In addition, assumption $\tuple{\neg o_{k^i+1}}$ restricts the sum to $k^i$. 
Notice that in the $(i+1)^{th}$ iteration, a new cardinality sum 
$\Psi(\mathbf{X}^{i+1},k^{i+1},b^{i+1})$ is added and earlier constraints are 
disabled as assumption $\tuple{\neg b^i}$ moves as unit clause $\insquareb{b^i}$.

\Omit{
Algorithm~\ref{alg:incremental-blocking} shows how the encoding of the 
cardinality constraint is used with the incremental blocking. 
Here, $\mathcal A$ and $\mathcal U$ denotes the set of assumptions and the 
set of unit clauses respectively.
For any iteration of a MaxSAT algorithm, the algorithm~\ref{alg:incremental-blocking}
starts by setting the previous assumptions to unit clauses. 
This disables all clauses of the previous cardinality constraint. 
Next, the algorithm creates a fresh blocking variable $b$, 
and sets $\neg b$ as an assumption. 
The cardinality constraint 
is encoded into CNF with the blocking variable $b$, i.e. the variable $b$ will 
be present in all clauses of the cardinality constraint. 
The sums from $k+1$ to $n$ are disabled for the current iteration by setting 
the output literals $\neg o_{k+1}, \ldots, \neg o_n$ as assumptions.
%
%
%
Algorithms~\ref{alg:linear-us} (Linear search Unsat-Sat) and~\ref{alg:msu3} (MSU3) 
from Section~\ref{sec:preliminaries} can be enhanced with 
incremental blocking by using Algorithm~\ref{alg:incremental-blocking} to 
encode the cardinality constraint to CNF. }
%

%
%
Assume the MaxSAT formula has a given optimum 
value $k_{opt}$. When considering Algorithm~\ref{alg:linear-us} and the Totalizer encoding, 
incremental blocking creates an encoding for each $k^i$ up to $k_{opt}$. Hence, 
the overall encoding  would have $O(\sum_{i=0}^{k_{opt}} ni) = O(nk_{opt}^2)$ auxiliary 
clauses. Though incremental blocking creates more clauses as compared to 
a non-incremental approach ($O(nk_{opt})$), keeping the inner state of the constraint 
solver across iterations significantly reduces the solving time. 
A similar reasoning can be made for Algorithm~\ref{alg:msu3} or any other MaxSAT algorithm 
that uses incremental blocking.

\Omit{
This contrasts with the non-incremental approach which only requires 
$O(nk)$ auxiliary clauses to encode the last cardinality constraint.
Incremental blocking keeps all learned clauses and the inner state of the solver, 
but has the downside of adding more clauses than the non-incremental approach.
}

\subsubsection{Fu-Malik Algorithm with Incremental Blocking.}

\DontPrintSemicolon
\SetKwFunction{soft}{soft}
\SetKwFunction{SAT}{SAT}
\SetKwFunction{decomposeSoft}{partitionSoft}
\SetKwFunction{first}{first}
\SetKwFunction{weight}{weight}
\SetKwFunction{encodeCNF}{CNF}
\SetKwFunction{min}{min}
\SetKwData{result}{satisfiable assignment to}
\SetKwData{unsat}{UNSAT}
\SetKwData{sat}{SAT}
\SetKwData{minc}{min$_\textnormal{c}$}
\SetKwData{true}{true}
\SetKwData{st}{st}
\SetVlineSkip{1pt}
\begin{algorithm}[!t]
  \small
  \KwIn{$\varphi = \varphi_h \cup \varphi_s$}
  \KwOut{satisfying assignment to $\varphi$}
  ($\mathcal \varphi_W, \varphi_{W_s}, \mathcal A, \mathcal B) \gets (\varphi, \varphi_s, \emptyset, \emptyset)$\;
  \While{\true}{
      $(\st, \nu, \varphi_C) \gets \SAT(\varphi, \mathcal A)$\;
        \If{$\st = \sat$}{
      \Return{$\nu$}\tcp*[r]{\footnotesize satisfying assignment to $\varphi$}
    }
      $V_R \gets \emptyset$\;
      \ForEach{$\omega \in (\varphi_C ~\cap~ \varphi_{W_s})$}{
        $V_R \gets V_R \cup \{ r \}$\tcp*[r]{\footnotesize r is a new relaxation variable}
        \highlight{$\omega_R \gets (\omega \setminus \mathcal B) \cup \{ r \} \cup \{ b \}$}\tcp*[r]{\footnotesize b is a new blocking variable}
        \highlight{$\mathcal B \gets \mathcal B \cup \{ b \}$}\;
        $\varphi_{W_s} \gets (\varphi_{W_s} \setminus \{ \omega \}) \cup \{ \omega_R\}$\;
        \highlight{$\mathcal A \gets (\mathcal A \setminus \{ \neg b' : b' \in \mathcal B \cap \omega\}) \cup \{ \neg b\}$}\tcp*[r]{\footnotesize enables $\omega_R$}
        \highlight{$\varphi_W \gets \varphi_W \cup \{ \omega_R \} \cup \{ b' : b' \in \mathcal B \cap \omega\ \}$}\tcp*[r]{\footnotesize disables $\omega$}
      }
      $\varphi_W \gets \varphi_W \cup \{ \encodeCNF( \sum_{r\in V_R} r \leq 1 ) \}$\;
    }
  \caption{Fu-Malik Algorithm with Incremental Blocking}\label{alg:fu-malik-inc}
\end{algorithm}

Incremental blocking can also be used for MaxSAT algorithms that do not 
update cardinality constraints but modify the formula at 
each iteration. For example, Fu-Malik algorithm (Algorithm~\ref{alg:fu-malik-orig}, Section~\ref{sec:preliminaries}) 
can be enhanced with incremental blocking. Algorithm~\ref{alg:fu-malik-inc} shows 
the modifications to Fu-Malik algorithm to support incremental blocking.
The main differences between the incremental and non-incremental versions of 
Fu-Malik algorithm are highlighted. 
%
For each soft clause $\omega$ in $\varphi_C$, Algorithm~\ref{alg:fu-malik-inc} copies 
$\omega$ into $\omega_R$ without blocking variables (line 10). Next, it adds a fresh blocking 
variable $b$ and a fresh relaxation variable $r$ to $\omega_R$ (line 10). 
The current soft clause $\omega_R$ is enabled by adding $\tuple{\neg b}$ to the set of assumptions, 
where $b$ is the blocking variable that occurs in $\omega_R$ (line 12). At the same time, 
the assumption $\tuple{\neg b'}$ is removed from the set of assumptions, where $b'$ is the blocking
variable that occurs in $\omega$ (line 12). Finally, the working formula $\varphi_W$ is updated with the 
new clause $\omega_R$, and with the unit clause $\insquareb{b'}$. Note that this unit clause 
disables $\omega$ from the working formula $\varphi_W$ since $\omega$ contains $b'$ and 
therefore is always satisfied.

\Omit{
For each soft clause $\omega$ in the 
unsatisfiable subformula, Algorithm~\ref{alg:fu-malik-inc} copies $\omega$ into $\omega_R$ without 
blocking variables (line 12). Next, it adds a fresh blocking variable $b$ and a 
fresh relaxation $r$ to $\omega_R$ (line 12). This step allows the previous soft clause to 
be replaced by a new soft clause. To disable $\omega$ the algorithm adds the 
blocking variable occurring in $\omega$ to the set of unit clauses (line 13). 
Next, the negation of the blocking variable in $\omega_R$ is added to the 
set of assumption variables to enable $\omega_R$ for the upcoming iterations. 
Finally, the formula $\varphi$ is updated with the new clause $\omega_R$. 
After all soft clauses have been relaxed, the at-most-one constraint is added 
over the relaxation variables as in the original algorithm. The difference is 
that now the algorithm also adds the set of unit clauses ($\mathcal U$) to 
 $\varphi$, which disables the soft clauses that occurred in $\varphi_C$ for upcoming 
 iterations (line 16).
}

The incremental version of Fu-Malik algorithm creates $m$ auxiliary clauses at each 
iteration, where $m$ is the number of soft clauses in the unsatisfiable subformula. 
However, the size of unsatisfiable subformulas tends to be small when compared 
to the total number of soft clauses. 
Note that the number of auxiliary clauses created by the incremental version 
of Fu-Malik is not as large as when incremental blocking is directly applied to 
cardinality encodings. 

\subsection{Incremental Weakening}
Since incremental blocking encodes a new cardinality constraint at each iteration,
this results in an increase in formula size at every iteration.
To circumvent this increase, one can build the cardinality 
sum only once, and incrementally weaken the cardinality bound ($k$).


%
%
\Omit{
\begin{definition}[Incremental Weakening]
Let $l_1 + \ldots + l_n \leq k$ be a cardinality constraint. Consider we have 
a CNF encoding of $l_1 + \ldots + l_n \leq k$ with output literals $o_1, \ldots, 
o_n$. To constrain sums higher than $k$, we add the output literals 
$\neg o_{k+1},$ $\ldots,$ $\neg o_n$ to the set of assumptions. On the the next 
iteration for a lower bound $k'$ higher than $k$, we do not need to rebuild the 
encoding but can instead update the assumptions with the output literals 
$\neg o_{k'+1}, \ldots, \neg o_n$. 
\end{definition}
}
Incremental weakening is similar to {\it incremental strengthening}~\cite{asin-constraints11}, but 
instead of constraining the output of the cardinality constraint with unit clauses 
it uses assumptions. Notice that incremental strengthening is used in linear 
search Sat-Unsat algorithms. In these algorithms, the cardinality bound 
decreases monotonically at each iteration. Therefore, the unit clauses that 
constrain the previous cardinality bound remain valid when considering the new 
bound. On the other hand, incremental weakening is used for MaxSAT algorithms 
that search on the lower bound of the optimal solution. For these algorithms, 
the restriction of the cardinality bound is only valid for the 
current iteration and must be updated for the upcoming iterations. 

\Omit{
Notice that incremental strengthening is used in linear 
search algorithms that search on the upper bound of the objective value of a 
MaxSAT instance. In these algorithms, the upper bound decreases monotonically at each 
iteration. Therefore, the unit clauses that constrain the previous upper bound 
remain valid when considering the new upper bound. 
On the other hand, incremental weakening is used for MaxSAT 
algorithms that search on the lower bound of the objective value of a MaxSAT 
instance. For these algorithms, the restriction of the lower bound is only valid for the 
current iteration and must be updated for the upcoming iterations. 
}

\begin{subequations}
\begin{align}
\Gamma(\mathbf{X},k) \equiv \mathtt{CNF}_{Tot^k}(\Sigma x_i) \label{eq:cardupper} \\
\varphi^i \equiv \varphi_W \cup  \Gamma(\mathbf{X},k_u) \cup \tuple{\neg o_{k^i+1},\dots,\neg o_{k_u}} \label{eq:incweak-i}\\
\varphi^{i+1} \equiv \varphi_W \cup  \Gamma(\mathbf{X},k_u) \cup \tuple{\neg o_{k^{i+1}+1},\dots,\neg o_{k_u}} \label{eq:incweak-inext}
\end{align}
\end{subequations}

Let $\Gamma(\mathbf{X},k)$ be the cardinality sum over input literals $x_1,\dots,x_n$ encoded
in CNF using Eq.~\ref{eq:totalizersum} and $k$-simplification. Then, for incremental weakening, at line~\ref{linearus:satcall} in Algorithm~\ref{alg:linear-us} and line~\ref{msu3:satcall} in
Algorithm~\ref{alg:msu3} we call the solver on $\varphi^i$ as defined in Eq.~\ref{eq:incweak-i} for the $i^{th}$ iteration. Note that $\Gamma(\mathbf{X},k_u)$ is encoded only once for 
a conservative upper bound $k_u$. For the $i^{th}$ iteration, we restrict the cardinality sum to $k^i$ using assumptions $\tuple{\neg o_{k^i+1},\dots,\neg o_{k_u}}$ (Eq.~\ref{eq:comparator}).
In the following iteration (Eq.~\ref{eq:incweak-inext}), we only change assumptions to restrict the cardinality sum to $k^{i+1}$.

To obtain a conservative upper bound $k_u$, we invoke the SAT solver over $\varphi_h$ 
to check if the set of hard clauses itself is satisfiable. If it is not
satisfiable, the original MaxSAT formula $\varphi$ can not be solved. 
However, if $\varphi_h$ is satisfiable, one can count the number of soft 
clauses that remain unsatisfied under the satisfying assignment for $\varphi_h$. 
This number can be used as $k_u$ since we know at least one assignment 
where $k_u$ many clauses remain unsatisfied. Therefore, the optimum 
value $k_{opt}$ must be smaller or equal to $k_u$.

\Omit{
Algorithm~\ref{alg:incremental-weak} shows how can the encoding of the cardinality 
constraint be used with the incremental weakening strategy. The algorithm checks 
if an encoding has already been built. If not, it 
creates the CNF encoding of the cardinality constraint. For any iteration of a 
MaxSAT algorithm, the sums from 
$k+1$ to $n$ are disabled by updating the set of assumptions with the 
output literals $\neg o_{k+1}, \ldots, \neg o_n$. 

We assume that the encoding allows the sum to go up to $n$, e.g. Totalizer encoding. 
This would lead 
%
to $O(n^2)$ auxiliary clauses since the $k$-simplification referred in Section~\ref{sec:cardinality} 
cannot not be used in this case. However, if we have an upper bound $k'$ on the value of the 
objective value of a MaxSAT instance, then we can use that upper bound to 
restrict our encoding for sums up to $k'+1$.
Such an upper bound can be obtained when we call the constraint solver over the 
hard clauses to check if the MaxSAT formula is unsatisfiable\footnote{Note that a 
model of the hard clauses can be used to check how many soft clauses are 
unsatisfied under that model. This gives an upper bound on the number of 
unsatisfied soft clauses.}. 
}
With an upper bound 
$k_u$, incremental weakening creates $O(nk_u)$ auxiliary clauses as opposed to 
$O(nk_{opt})$ of the non-incremental approach. However, a non-incremental approach
builds a new formula of size $O(nk_{opt})$ for every iteration, whereas incremental weakening builds the formula only once
keeping the internal state and learned clauses across iterations. This results in a significant performance boost for MaxSAT algorithms using incremental weakening.
%

Incremental weakening does not allow 
the number of input literals in the cardinality constraint to change. Therefore, it 
does not directly support the MSU3 Algorithm (Algorithm~\ref{alg:msu3}, Section~\ref{sec:preliminaries}). 
To use incremental weakening with Algorithm~\ref{alg:msu3}, we modify the 
algorithm to relax all soft clauses and build a cardinality constraint over 
all relaxation variables. The relaxation variables $r_i$ that do not appear 
in an unsatisfiable subformula $\varphi_C$ are added as assumptions of the 
form $\tuple{\neg r_i}$.
This enforces the soft clauses corresponding to the relaxation variables until 
these clauses occur in $\varphi_C$.
When they do occur, assumptions $\neg r_i$ are removed 
and their value is now only restricted by the cardinality 
constraint.
Even though this procedure allows the incremental 
weakening approach to be used with Algorithm~\ref{alg:msu3}, it does not benefit from smaller 
encodings resulting from having less input literals in the cardinality constraint. Therefore, 
the non-incremental approach may create a much smaller encoding than the 
incremental weakening approach for Algorithm~\ref{alg:msu3}.

\subsection{Iterative Encoding}
\begin{figure}[t]
\centering
\begin{tikzpicture}[level/.style={sibling distance=50mm/#1},scale=0.8, every node/.style={scale=0.8}]
\node (a) {$(A:a_1,a_2 \rightarrow a_1, a_2, a_3,a_4 : 2 \rightarrow 4)$}
	child { node (b) {$(B:b_1,b_2:2)$ } 
		child { node (d) {$(D:l_1:1)$} } 
		child { node (e) {$(E:l_2:1)$} } }
	child { node (c) {$(C:c_1, c_2, \rightarrow c_1, c_2, c_3 : 2 \rightarrow 3)$} 
                child { node (g) { $(G: l_3:1)$}}
		child { node (f) {$(F:f_1, f_2:2)$} 
			child { node (h) {$(H:l_4:1)$}}
			child { node (i) {$(I:l_5:1)$}}} };
\node (j) [right of = a,node distance=8cm]{$(J:j_1,j_2:2)$}
	child { node [right=3mm] (k) {$(K:l_7:1)$} }
	child { node  [left=3mm] (l) {$(L:l_8:1)$}};
\node (p) [above right = 0.5cm and 0.6cm of a] {$(O:o_1,\dots,o_4:4)$};
\draw [dashed] (p) -- (j);
\draw [dashed] (p) -- (a);

\end{tikzpicture}
\caption{Transforming $l_1 + \dots + l_5 \leq 1$ and $l_7+l_8 \leq 1$ into 
$l_1 + \ldots + l_5 + l_7 + l_8 \leq 3$}
\label{fig:mergetree}
\end{figure}
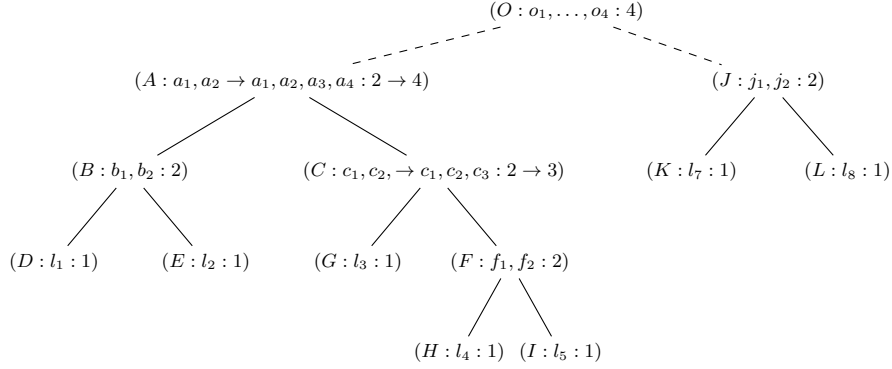

\Omit{
\begin{itemize}
\item Iterative strategy
\item Cardinality encodings properties:
\begin{itemize}
\item The clauses that encode each {\it k} are disjoint
\item Two disjoint cardinality constraints can be merged into one iff: 
(i) the literals occurring in each cardinality constraint are disjoint, and 
(ii) the {\it k} value for the new cardinality constraint is larger or equal than the previous {\it k'} values.
\item Cardinality constraints with the above properties may be built in an iterative fashion (e.g. sequential, totalizer, modulo totalizer)
\item Use the totalizer encoding to show how to build an encoding using an iterative strategy
\end{itemize}
\end{itemize}
}

Incremental weakening uses a conservative upper bound (e.g., $k_u$) on the number of unsatisfied soft clauses 
in order to encode the cardinality constraint only once.
However, this upper bound may be much larger than the optimum value (e.g. $k_{opt}$) which may result in a larger 
encoding than the non-incremental approach.
In addition, incremental weakening does not allow  the set of input literals in the cardinality constraint to change.
Therefore, MaxSAT algorithms that increase the input literals of the cardinality constraint can not take advantage of incremental weakening.
\Omit{
Moreover, incremental weakening does not take advantage of MaxSAT algorithms that 
increase the number of input literals of cardinality constraints during the course of the search. 
This may also result in a much larger encoding than the corresponding non-incremental approach.}
To remedy this situation, we propose to encode the cardinality constraint in an iterative fashion. At each iteration
of the MaxSAT algorithm, the encoding of the cardinality constraint is augmented with clauses that allow the sum
of input literals to go up to $k$ for the current iteration. We call this approach \emph{iterative encoding}.
\Omit{
Instead of building an encoding that allows the sum to go up to a given upper bound, one can 
build the encoding in an iterative fashion, i.e. at each iteration of the MaxSAT algorithm 
only the clauses that encode the current sum are encoded into CNF. }
%

\Omit{
\begin{definition}[$k$-Disjointness of a Cardinality Encoding]
Let $l_1 + \ldots + l_n \leq k$ be a cardinality constraint. 
Consider a succession of possible $k$ values, such that $k_1 < k_2 < \ldots < k_i$.
A cardinality encoding has the property of \emph{$k$-disjointness} if the clauses necessary to encode 
$k_j$ are disjoint from the clauses necessary to encode $k_{j+1}$, with $1 \leq j \leq i$. 
\end{definition}

In section~\ref{sec:cardinality} we referred that the Totalizer encoding can be 
reduced in the number of auxiliary clause by using the $k$-simplification technique. 
This technique is based on the $k$-disjointness property of the Totalizer encoding. 
The linear search Unsat-Sat algorithm (Algorithm~\ref{alg:linear-us}, section~\ref{sec:preliminaries})
 can be enhanced with incrementality by using an 
iterative version of the Totalizer encoding that exploits the $k$-Disjointness 
property.

\begin{definition}[$n$-Disjointness of a Cardinality Encoding]
Let $x_1 + \ldots + x_{n_1} \leq k_1$ and $l_1 + \ldots + l_{n_2} \leq k_2$ be two cardinality constraints. 
Consider that the literals that occur in each cardinality constraint are disjoint.
A cardinality encoding has the \emph{$n$-disjointness} property if it can encode each of the cardinality constraints 
independently and then merge both encodings to encode $x_1 + \ldots + x_{n_1} + l_1 + \ldots + l_{n_2} \leq 
k_3$, with $k_3 \geq k_1$ and $k_3 \geq k_2$.
\end{definition}
}

Let us take a look at Fig.~\ref{fig:mergetree} to see how iterative encoding proceeds. Assume that for a particular iteration,
we needed to encode $l_1+\dots+l_5 \leq 1$. This can be accomplished using the subtree rooted at $A$. Since the bound for this iteration is $k=1$, we only need $k+1=2$,
$node\_vars$ at every node as described in $k$-simplification in Section~\ref{sec:cardinality}. In the next iteration, suppose we need to encode $l_1+\dots+l_5+l_7+l_8 \leq 3$.
Observation~\ref{obs:monotonic} allows us to augment the formula for subtree rooted at $A$ to allow $l_1+\dots+l_5$ to sum up to $4$. 
This is done by increasing the output variables of node $A$ to sum up to $4$ and adding the respective clauses that encode sums $3$ and $4$. Similarly, for node $C$ the output variables are increased to sum up to $3$ and the clauses that sum up to $3$ are added to the formula. 
For the additional input literals $l_7$ and $l_8$ we encode
the subtree rooted at $J$. Observation~\ref{obs:independent} allows us to merge trees rooted at $A$ and $J$ by creating a new parent node $O$ which sums up to $4$ since $A$ and $J$ have disjoint
sets of input literals. To restrict the number of input literals being set to $true$ to $3$, we only need to add $\neg o_4$ as described in Eq.~\ref{eq:comparator}.

In general, if the cardinality constraint changes from $x_1+\dots+x_n \leq k_1$ ($k_1<n$) to $x_1+\dots+x_n+y_1+\dots+y_m \leq k_2$ where $k_1\leq k_2$ then we do the following : (1) Remove the 
assumption over output literal $\neg o_{k_1+1}$ which restricts the sum of $x_1\dots,x_n$ to $k_1$. (2) Augment the formula for $x_1,\dots,x_n$ to sum up to $min(k_2+1,n)$. (3) Encode the formula
over $y_1,\dots,y_m$ to sum up to $min(k_2+1,m)$. (4) Conjunct these two formulas and augment the resulting formula using Eq.~\ref{eq:totalizersum} and $k$-simplification in order to encode
$x_1+\dots+x_n+y_1+\dots+y_m \leq k_2$. Since iterative encoding always adds clauses to the existing formula and changes assumptions, it allows us to retain the internal state of the solver across iterations.

Linear search Unsat-Sat algorithm (Algorithm~\ref{alg:linear-us}, Section~\ref{sec:preliminaries}) increases the 
cardinality bound by 1 at each iteration but does not change the set of input literals of the cardinality constraint. 
Therefore, to apply iterative encoding to this algorithm we only perform steps (1) and (2).
On the other hand, MSU3 algorithm (Algorithm~\ref{alg:msu3}, Section~\ref{sec:preliminaries}) 
may change the set of input literals of the cardinality constraint between iterations. 
Therefore, iterative encoding is applied to MSU3 by performing steps (1) to (4).

Since at every iteration, bare minimum number of clauses necessary to encode the cardinality constraint for that iteration is added, the size of the encoding remains small throughout the run of the MaxSAT algorithm. Iterative encoding is not only faster but allows us to solve more problem instances as compared to non-incremental approaches.

\Omit{
\begin{figure}
\centering
\begin{tikzpicture}[level/.style={sibling distance=50mm/#1},scale=0.8, every node/.style={scale=0.8}]
\node (a) {$(A:a_1,a_2,a_3,a_4,a_5:5)$}
	child { node (b) {$(B:b_1,b_2:2)$ } 
		child { node (d) {$(D:x_1:1)$} } 
		child { node (e) {$(E:x_2:1)$} } }
	child { node (c) {$(C:c_1, c_2, c_3:3)$} 
                child { node (g) { $(G: x_3:1)$}}
		child { node (f) {$(F:f_1, f_2:2)$} 
			child { node (h) {$(H:x_4:1)$}}
			child { node (i) {$(I:x_5:1)$}}} };
\node (j) [right of = a,node distance=8cm]{$(J:j_1,j_2:2)$}
	child { node [right=3mm] (k) {$(K:x_7:1)$} }
	child { node  [left=3mm] (l) {$(L:x_8:1)$}};
\node (p) [above right = 0.5cm and 0.6cm of a] {$(P:p_1,\dots,p_7:7)$};
\draw [dashed] (p) -- (j);
\draw [dashed] (p) -- (a);

\end{tikzpicture}
\caption{Merging of  $x_1 + \dots + x_5 \leq k_1$ and $x_7+x_8 \leq k_2$}
\label{fig:mergetree}
\end{figure}
}


\section{Related Work}
\label{sec:related}
The first use of incremental SAT solving can be traced back to the 90's with the seminal work of John Hooker~\cite{hooker-jlp93}. Initially, only a subset of constraints is considered. At each iteration, more constraints are added to the formula.
Later, incremental approaches were adopted by constraint solvers in the context of SAT~\cite{whittemore-dac01,minisat-sat03} and SAT extensions~\cite{qmaxsat-jsat12,audemard-sat13}. 
%

Assumptions are widely used for incremental SAT~\cite{minisat-assumptions,nadel-sat12}. The minisat solver~\cite{minisat-sat03} interface allows the definition of a set of assumptions. Alternatively, the interface of zchaff~\cite{mahajan-sat04} allows removing groups of clauses.


Although not implemented, the work of Fu and Malik in MaxSAT~\cite{FM06} discusses how learned clauses may be kept from one SAT iteration to the next one. In Pseudo Boolean Optimization (PBO), early implementations include the use of incremental strengthening in minisat+~\cite{een-jsat06}. Linear search Sat-Unsat algorithms~\cite{qmaxsat-jsat12,berre-jsat10} are implemented incrementally. A critical issue is on keeping {\em safe} learned clauses in successive iterations of a core-guided algorithm~\cite{martins-ai12}. Quantified Boolean Formula (QBF) solving has successfully been made incremental~\cite{Lonsing-corr14} and further applied to verification~\cite{marin-date12}.

In the context of SAT, incremental approaches exist for building encodings and identifying Minimal Unsatisfiable Subformulas (MUSes). For example, an incremental translation to CNF uses unit clauses to simplify the pseudo-Boolean constraint before translating it to CNF~\cite{manolios-fmcad11}. More recent work {lazily} decomposes complex constraints into a set of clauses~\cite{abio-cp12}. The identification of MUSes has been made incremental by Liffton {\em et al.}~\cite{liffiton-jar08}. Later on, the SAT solver Glucose has been made incremental using assumptions and applied to MUS extraction~\cite{audemard-sat13}.

Incrementality is also present in other SAT-related domains such as Satisfiability Modulo Theories (SMT) and Bounded Model Checking (BMC). 
The SMT-LIB v2.0~\cite{BarST-RR-10} defines the operations {\em push} and {\em pop} to work with a stack containing a set of formulas to be jointly solved. The MaxSAT solvers WPM1 and WPM2~\cite{ansotegui-cp13} use the SMT solver Yices~\cite{yices} which supports incrementality. Its use resembles the blocking strategy. 
%
The use of SAT solvers in BMC is known to benefit from incrementality, either by implementing incremental SAT solving~\cite{strichman-charme01} or by using assumptions~\cite{minisat-assumptions}.

In the context of Constraint Satisfaction Problems (CSPs), incremental formulations, incremental propagation and incremental solving are worth mentioning. 
Incrementality is naturally present in Dynamic CSPs (DCSPs)~\cite{dechter-aaai88}. In DCSPs, the formulation of a problem evolves over time by adding and/or removing variables and constraints. {\em Nogoods} can eventually be carried from one formulation to the next one. DCSPs make use of an incremental arc consistency algorithm~\cite{debruyne-ictai96}.
Incremental propagation in CSP~\cite{lagerkvist-cp07,cheng-ecai06} makes use of {\em advisors} which give propagators a detailed view of the dynamic changes between propagator runs. Advisors enable the implementation of optimal algorithms for important constraints.
Search in CSP is inherently incremental. From the first implementations, the approach to solve many CSPs is to incrementally build a solution, backtracking when an infeasibility is detected, until a solution is found or the problem is proven to have no solution~\cite{vanbeek-hcp06}. More recently, incrementality has been implemented in global constraints mostly due to efficiency reasons~\cite{vanhoeve-hcp06}.


\section{Experimental Results}
\label{sec:results}
We used all partial MaxSAT instances (627) from the industrial 
category of the MaxSAT Evaluation 2013\footnote{Benchmarks available at 
\url{http://maxsat.ia.udl.cat/13/benchmarks/}} as a benchmark for our experiments. The evaluation was performed on two 
AMD Opteron 6276 processors (2.3 GHz) running Fedora 18	with a timeout of 
1,800 seconds and a memory limit of 8 GB. 
We implemented all algorithms described in section~\ref{sec:preliminaries} (Linear search Unsat-Sat, 
Fu-Malik, and M\-SU3), as well as their incremental counterparts 
on top of {\sc open-wbo}~\cite{martins-sat14}. 
{\sc open-wbo} is a modular open source MaxSAT solver that is easy to modify and is 
competitive with state-of-the-art MaxSAT solvers\footnote{{\sc open-wbo} with iterative encoding achieved
first place for unweighted MaxSAT industrial and second place for partial MaxSAT industrial category in
MaxSAT 2014 evaluations. \url{http://maxsat.ia.udl.cat/results/}}.

Table~\ref{tbl:results} shows the number of instances solved ({\it \#Inst}) by the described MaxSAT 
algorithms using the different approaches, namely, non-incremental approach ({\it none}), 
incremental blocking ({\it blocking}), incremental weakening ({\it weakening}), 
and iterative encoding ({\it iterative}). 
Table~\ref{tbl:results} also shows the median speedup\footnote{The speedup of an instance is 
measured as the ratio of the solving time of the non-incremental approach to the solving time of the respective 
incremental approach.} for instances that have been solved by all incremental approaches for a given algorithm. 

\begin{table}[!t]
\caption{Number of instances solved by the different incremental approaches and median speedup of solved instances}
\label{tbl:results}
\begin{center}
\begin{tabular}{l | rr | rr | rr | rr|}
\cline{2-9}
& \multicolumn{2}{c|}{None} & \multicolumn{2}{c|}{Blocking} & \multicolumn{2}{c|}{Weakening} & \multicolumn{2}{c|}{Iterative}\\
\cline{2-9}
& \#Inst & Speedup & \#Inst & Speedup & \#Inst & Speedup & \#Inst  & Speedup\\
\hline
\multicolumn{1}{|l|}{Fu-Malik} & 366 & 1.0 & \bf 388 & \bf 2.4 & - & - & - & -\\
\multicolumn{1}{|l|}{LinearUS} & 477 & 1.0 & 446 & 1.6 & \bf 498 & \bf 2.3 & \bf 509 & \bf 2.4\\
\multicolumn{1}{|l|}{MSU3} &  517 & 1.0 & 488 & 1.6 & 504 & 2.0 & \bf 541 & \bf 3.6\\
\hline
\end{tabular}
\end{center}
\end{table}

Fu-Malik with incremental blocking significantly outperforms the non-incremental 
algorithm. Incremental blocking not only solves more instances but also is significantly faster 
than the non-incremental algorithm. From those instances which were solved by both approaches,  
50\% of them have a speedup of at least 2.4.
%
Incremental weakening and iterative encoding cannot be used with 
the Fu-Malik algorithm since it only uses at most one constraints 
and modifies the 
formula across iterations of the algorithm.

Linear search Unsat-Sat ({LinearUS}) with incremental blocking solves 
less instances than the non-incremental approach. 
Incremental blocking encodes a new cardinality constraint at each iteration of 
the MaxSAT algorithm, causing the formula to grow too large resulting in termination due to memory outs.
However, for those instances that were solved successfully, incremental blocking 
was 60\% faster than the original LinearUS.
Incremental weakening allows MaxSAT algorithms to solve more instances with 
significant speedup. Since the cardinality constraint is encoded only once, 
the size of the formula remains almost constant across iterations. The majority 
of the instances are solved at least 2$\times$ faster. Iterative encoding 
outperforms all other approaches. Smaller formula sizes due to iterative encoding
allows it to solve more instances as compared to incremental weakening. 

MSU3 with incremental blocking solves less instances as compared to the original MSU3
but it is faster for instances solved by both approaches. 
Similar results have been observed for the LinearUS algorithm with incremental blocking.
%
Incremental weakening outperforms incremental blocking in the number of solved 
instances as well as in terms of solving time. However, 
incremental weakening solves less instances than the non-incremental approach,
since incremental weakening is not flexible to directly support the increase in the 
number of input literals of the cardinality constraint. A non-incremental approach 
may need to impose the cardinality constraint over a small subset of relaxation variables.
Incremental weakening does not enjoy this benefit due to its inflexibility. This may result in
incremental weakening producing a larger encoding for certain problem instances.
Iterative encoding solves more instances and is significantly faster than the 
non-incremental approach. Iterative encoding only encodes the clauses that 
are needed at each iteration of the MaxSAT algorithm, allowing for an 
encoding with a similar size to the non-incremental approach. Most instances 
are solved at least 3.6$\times$ faster with iterative encoding than without it.

\begin{figure}[!t] 
  \begin{subfigure}[b]{0.5\linewidth}
    \centering
    \includegraphics[width=0.75\linewidth]{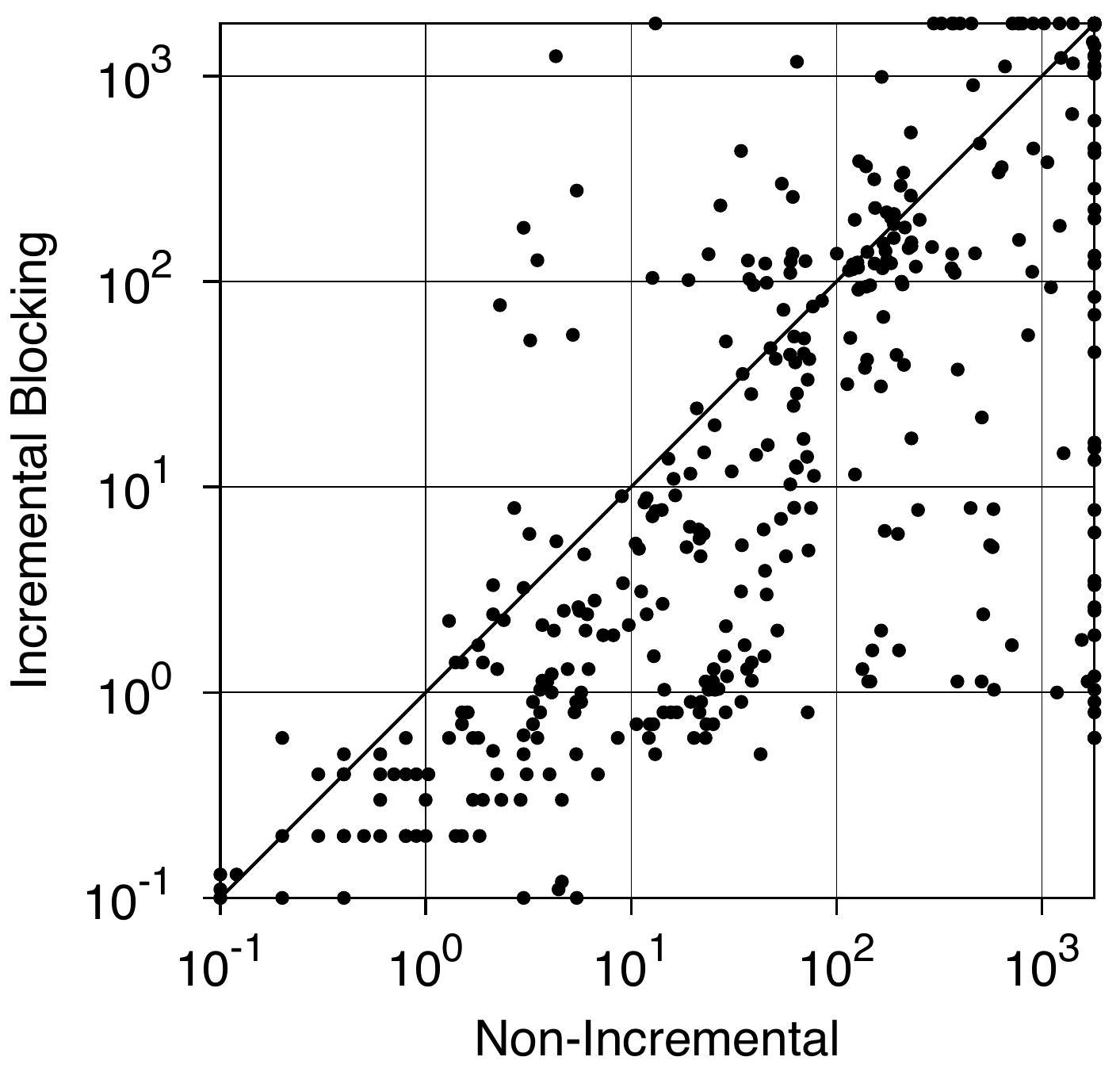} 
    \caption{Fu-Malik Algorithm:\\ Non-Incremental vs. Incremental Blocking} 
    \label{fig:blocking} 
    \vspace{4ex}
    \hfill
  \end{subfigure}
  \begin{subfigure}[b]{0.5\linewidth}
    \centering
    \includegraphics[width=0.75\linewidth]{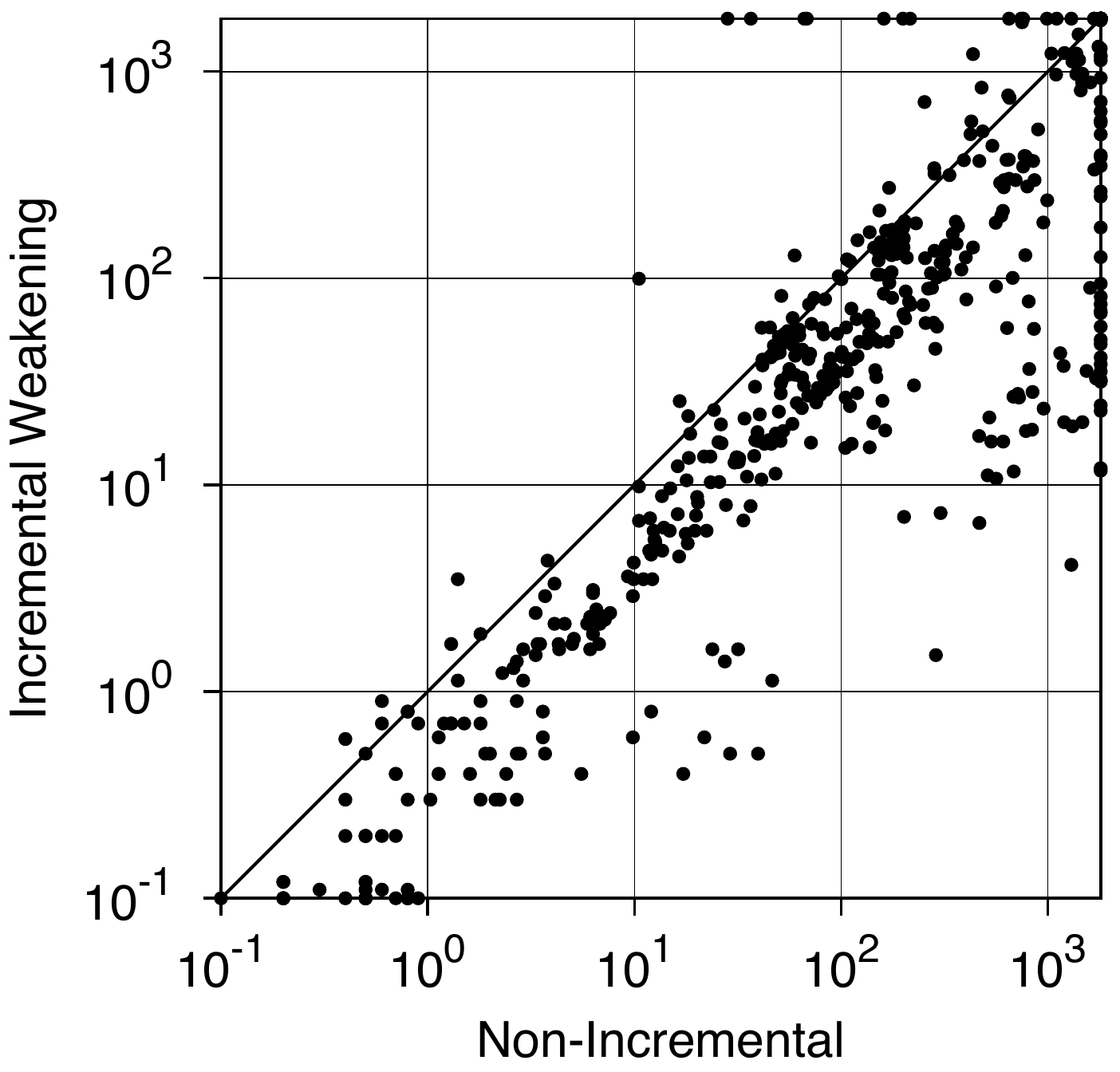} 
    \caption{LinearUS Algorithm:\\ Non-Incremental vs. Incremental Weakening} 
    \label{fig:weakening} 
    \vspace{4ex}
    \hfill
  \end{subfigure} 
  \begin{subfigure}[b]{0.5\linewidth}
    \centering
    \includegraphics[width=0.75\linewidth]{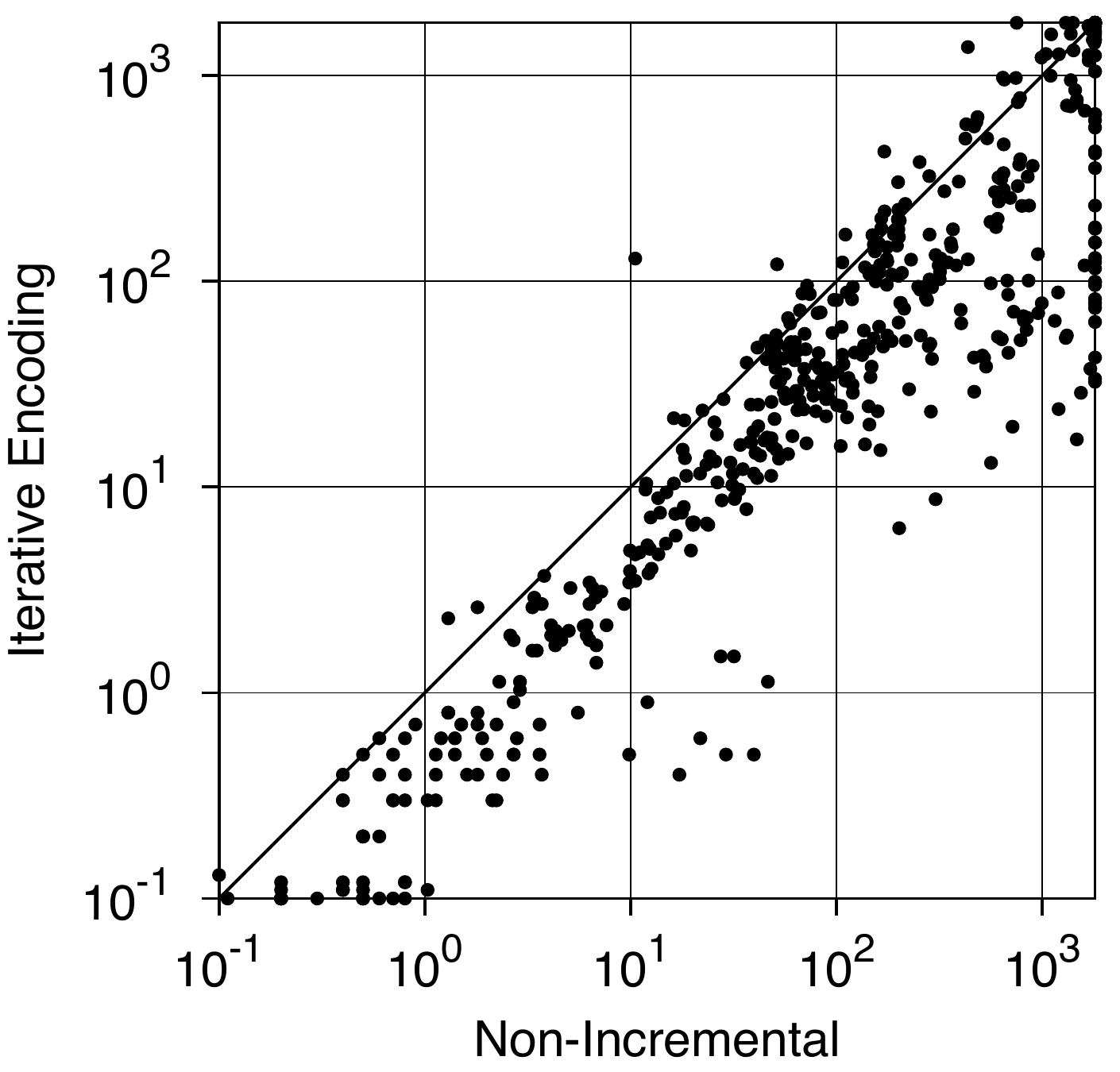} 
    \caption{LinearUS Algorithm:\\ Non-Incremental vs. Iterative Encoding} 
    \label{fig:linear-us-iterative} 
  \end{subfigure}
  \begin{subfigure}[b]{0.5\linewidth}
    \centering
    \includegraphics[width=0.75\linewidth]{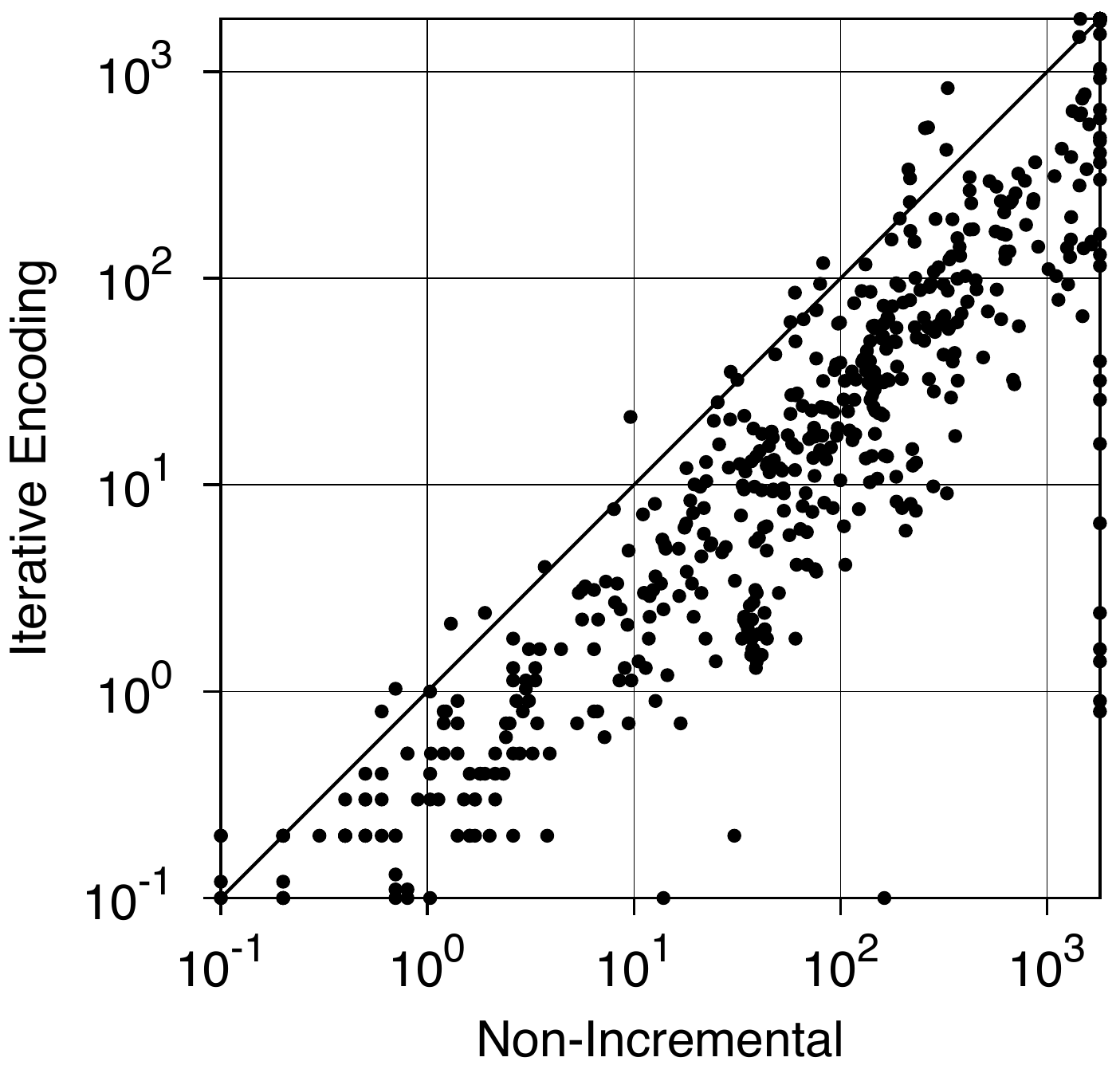} 
    \caption{MSU3 Algorithm:\\ Non-Incremental vs. Iterative Encoding} 
    \label{fig:msu3-iterative} 
  \end{subfigure} 
  \caption{Impact of incremental approaches}
  \label{fig:incremental} 
\end{figure}


Fig.~\ref{fig:incremental} shows scatter plots that compare the non-incremental and 
incremental ap\-proaches which are highlighted in Table~\ref{tbl:results}. 
Each point in the plot corresponds 
to a problem instance, where the x-axis corresponds to the run time required 
by non-incremental approaches and the y-axis corresponds to the run time required by 
incremental approaches. Instances that are above the diagonal are solved faster 
when using a non-incremental approach, whereas instances that are below the 
diagonal are solved faster when using an incremental approach. Incremental approaches that we propose in this paper
clearly assert their dominance over their non-incremental counterparts integrated with all three algorithms as shown in Fig.~
\ref{fig:incremental}.  This is particularly 
evident in the MSU3 algorithm where the majority of the instances are solved 
much faster with iterative encoding. For example, for 30\% of the instances solved by MSU3 with and without iterative encoding, 
iterative encoding is at least 6$\times$ faster than 
the non-incremental approach. For 10\% of the instances solved by both approaches, 
iterative encoding boosts MSU3 with at least 14$\times$ speedup.

\begin{figure}[ht]
\centering
    \includegraphics[height=6.8cm,width=10.6cm]{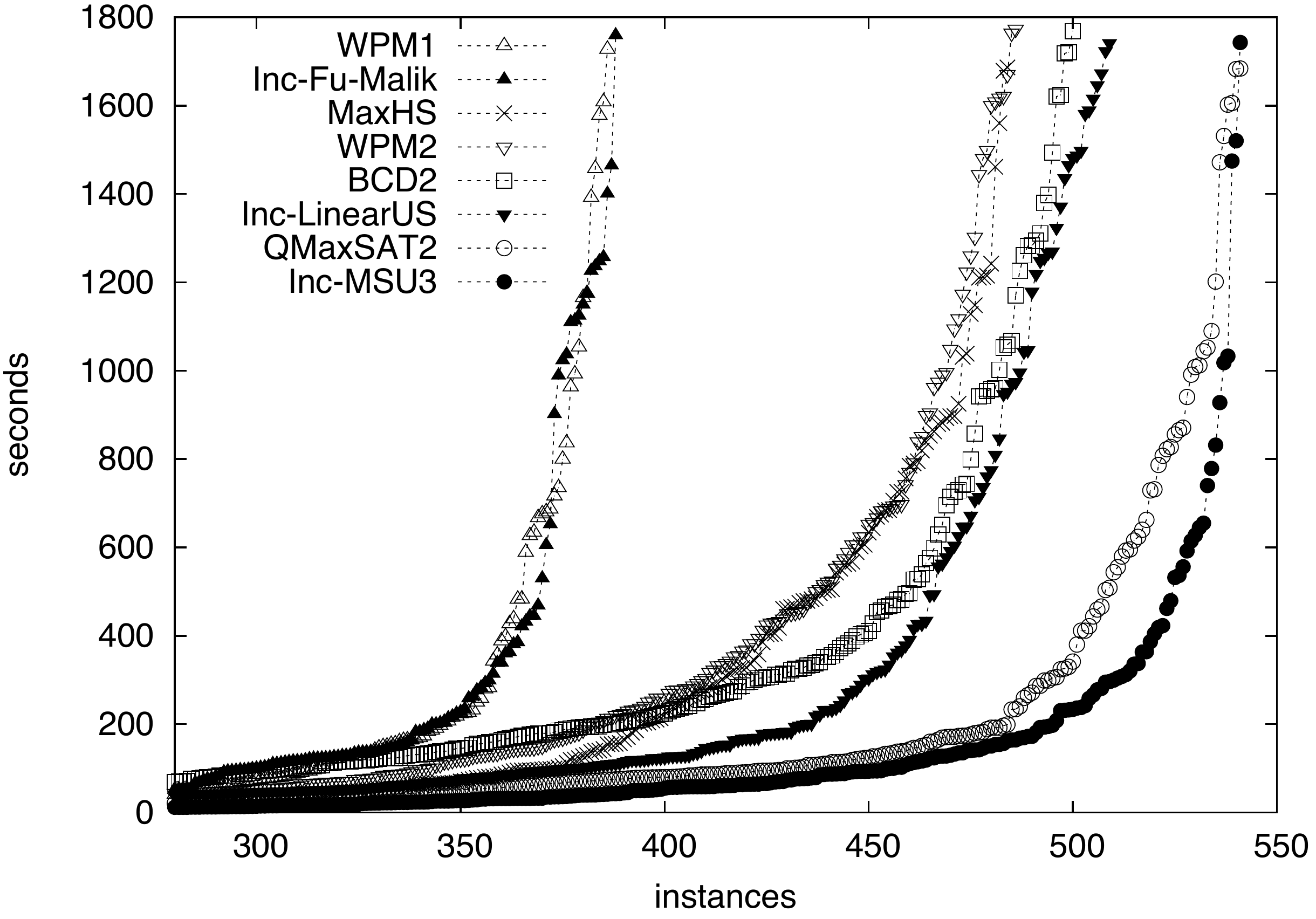} 
    \caption{Running times of state-of-the-art MaxSAT solvers} 
    \label{fig:cactus} 
\end{figure}

Fig.~\ref{fig:cactus} shows a cactus plot with the running times of 
state-of-the-art MaxSAT solvers used in the MaxSAT Evaluation 2013\footnote{Only 
single engine solvers have been considered in this evaluation, therefore we did not include ISAC+ (a portfolio MaxSAT solver)~\cite{Kadioglu12}.} 
(WPM1~\cite{ansotegui-sat09}, WPM2~\cite{ansotegui-aaai10,ansotegui-cp13}, MaxHS~\cite{davies-sat13,davies-cp13}, 
BCD2~\cite{morgado-sat12}, QMaxSAT2~\cite{qmaxsat-jsat12}) and 
the best incremental algorithms presented in this paper (incremental blocking 
Fu-Malik, iterative encoding LinearUS and MSU3).

Fu-Malik and WPM1 use similar MaxSAT algorithms. Moreover, WPM1 has a similar incremental
strategy due to the incremental SMT solver that is used by WPM1.
Since both solvers used similar techniques, it is not surprising that their performance is similar.
Even though LinearUS uses a simple MaxSAT algorithm, it is 
competitive with more complex state-of-the-art MaxSAT algorithms. This is mostly 
due to the incremental approach that is being used in LinearUS and shows the 
importance of using an efficient incremental approach.
MSU3 and QMaxSAT perform complementary searches but both use incrementality and 
have similar performances.
Iterative encoding is not restricted to MSU3 and may be used in other MaxSAT 
algorithms, such as WPM2 and BCD2. It is expected that if those algorithms are 
enhanced with the incremental iterative encoding, their performance might rise 
to values similar or higher than those of QMaxSAT and MSU3.

\section{Conclusions and Future Work}
\label{sec:conclusions}

Several state of the art MaxSAT algorithms are based on solving a sequence
of closely related SAT formulas. However, although incrementality is not
a new technique, it is seldom used in MaxSAT algorithms that search on
the lower bound of the optimum solution.
In this paper, we describe and propose new techniques to incrementally
modify cardinality constraints used in several MaxSAT algorithms, namely 
in linear Unsat-Sat search, the classic Fu-Malik algorithm and MSU3 
core-guided algorithm.

Experimental results show the effectiveness of the techniques
proposed in the paper. The incremental versions of the MaxSAT
algorithms clearly outperform the non-incremental versions, both 
in terms of speed and number of solved instances. Furthermore, 
the proposed techniques can be integrated in other core-guided
algorithms such as WPM2 and BCD2, among others.

Finally, the paper also describes that in general it is possible to 
perform iterative encoding of cardinality constraints using the 
Totalizer encoding. Therefore, the use of this technique is not
limited to the scope of MaxSAT algorithms. As future work, we propose
to integrate these techniques in other domains where cardinality
constraints are used, and to extend incrementality to other effective
cardinality constraints encodings.



\bibliography{cp14}

\begin{thebibliography}{10}
\providecommand{\url}[1]{\texttt{#1}}
\providecommand{\urlprefix}{URL }

\bibitem{abio-cp12}
Ab\'{\i}o, I., Stuckey, P.J.: {Conflict Directed Lazy Decomposition}. In:
  Milano, M. (ed.) Principles and Practice of Constraint Programming. LNCS,
  vol. 7514, pp. 70--85. Springer (2012)

\bibitem{ansotegui-cp13}
Ans{\'o}tegui, C., Bonet, M.L., Gab{\`a}s, J., Levy, J.: {Improving WPM2 for
  (Weighted) Partial MaxSAT}. In: Principles and Practice of Constraint
  Programming. LNCS, vol. 8124, pp. 117--132. Springer (2013)

\bibitem{ansotegui-sat09}
Ans{\'o}tegui, C., Bonet, M.L., Levy, J.: {Solving (Weighted) Partial MaxSAT
  through Satisfiability Testing}. In: Kullmann  \cite{DBLP:conf/sat/2009}, pp.
  427--440

\bibitem{ansotegui-aaai10}
Ans{\'o}tegui, C., Bonet, M.L., Levy, J.: {A New Algorithm for Weighted Partial
  MaxSAT}. In: Fox, M., Poole, D. (eds.) AAAI Conference on Artificial
  Intelligence. AAAI Press (2010)

\bibitem{argelich10}
Argelich, J., Berre, D.L., Lynce, I., Marques-Silva, J., Rapicault, P.:
  {Solving Linux Upgradeability Problems Using Boolean Optimization}. In:
  Workshop on Logics for Component Configuration. pp. 11--22 (2010)

\bibitem{asin12}
As{\'i}n, R., Nieuwenhuis, R.: {Curriculum-based course timetabling with SAT
  and MaxSAT}. Annals of Operations Research pp. 1--21 (2012)

\bibitem{asin-constraints11}
As\'{\i}n, R., Nieuwenhuis, R., Oliveras, A., Rodr\'{\i}guez-Carbonell, E.:
  {Cardinality Networks: a theoretical and empirical study}. Constraints
  16(2),  195--221 (2011)

\bibitem{audemard-sat13}
Audemard, G., Lagniez, J.M., Simon, L.: {Improving Glucose for Incremental SAT
  Solving with Assumptions: Application to MUS Extraction}. In: J{\"a}rvisalo,
  M., Gelder, A.V. (eds.) International Conference on Theory and Applications
  of Satisfiability Testing. LNCS, vol. 7962, pp. 309--317. Springer (2013)

\bibitem{bailleux-cp03}
Bailleux, O., Boufkhad, Y.: {Efficient CNF Encoding of Boolean Cardinality
  Constraints}. In: Rossi, F. (ed.) Principles and Practice of Constraint
  Programming. LNCS, vol. 2833, pp. 108--122. Springer (2003)

\bibitem{BarST-RR-10}
Barrett, C., Stump, A., Tinelli, C.: {The SMT-LIB Standard: Version 2.0}. Tech.
  rep., Department of Computer Science, The University of Iowa (2010),
  available at {\tt www.SMT-LIB.org}

\bibitem{buttner-icaps05}
B{\"u}ttner, M., Rintanen, J.: {Satisfiability Planning with Constraints on the
  Number of Actions}. In: Biundo, S., Myers, K.L., Rajan, K. (eds.)
  International Conference on Automated Planning and Scheduling. pp. 292--299.
  AAAI (2005)

\bibitem{chen10}
Chen, Y., Safarpour, S., Marques-Silva, J., Veneris, A.G.: {Automated Design
  Debugging With Maximum Satisfiability}. IEEE Transactions on CAD of
  Integrated Circuits and Systems  29(11),  1804--1817 (2010)

\bibitem{cheng-ecai06}
Cheng, K.C.K., Yap, R.H.C.: {Maintaining Generalized Arc Consistency on Ad-Hoc
  n-Ary Boolean Constraints}. In: Brewka, G., Coradeschi, S., Perini, A.,
  Traverso, P. (eds.) European Conference on Artificial Intelligence. Frontiers
  in Artificial Intelligence and Applications, vol. 141, pp. 78--82. IOS Press
  (2006)

\bibitem{DBLP:conf/sat/2012}
Cimatti, A., Sebastiani, R. (eds.): Theory and Applications of Satisfiability
  Testing - SAT 2012 - 15th International Conference, Trento, Italy, June
  17-20, 2012. Proceedings, LNCS, vol. 7317. Springer (2012)

\bibitem{davies-sat13}
Davies, J., Bacchus, F.: {Exploiting the Power of mip Solvers in maxsat}. In:
  J{\"a}rvisalo, M., Gelder, A.V. (eds.) International Conference on Theory and
  Applications of Satisfiability Testing. LNCS, vol. 7962, pp. 166--181.
  Springer (2013)

\bibitem{davies-cp13}
Davies, J., Bacchus, F.: {Postponing Optimization to Speed Up MAXSAT Solving}.
  In: Schulte, C. (ed.) Principles and Practice of Constraint Programming.
  LNCS, vol. 8124, pp. 247--262. Springer (2013)

\bibitem{debruyne-ictai96}
Debruyne, R.: {Arc-Consistency in Dynamic CSPs Is No More Prohibitive}. In:
  International Conference on Tools with Artificial Intelligence. pp. 299--307.
  IEEE (1996)

\bibitem{dechter-aaai88}
Dechter, R., Dechter, A.: {Belief Maintenance in Dynamic Constraint Networks}.
  In: Shrobe, H.E., Mitchell, T.M., Smith, R.G. (eds.) AAAI Conference on
  Artificial Intelligence. pp. 37--42. AAAI Press / The MIT Press (1988)

\bibitem{yices}
Dutertre, B., de~Moura, L.M.: {A Fast Linear-Arithmetic Solver for DPLL(T)}.
  In: Ball, T., Jones, R.B. (eds.) Computer Aided Verification. LNCS, vol.
  4144, pp. 81--94. Springer (2006)

\bibitem{een-jsat06}
E\'en, N., S\"orensson, N.: {Translating Pseudo-{B}oolean Constraints into
  {SAT}}. Journal on Satisfiability, Boolean Modeling and Computation  2,
  1--26 (2006)

\bibitem{minisat-sat03}
E{\'e}n, N., S{\"o}rensson, N.: {An Extensible {SAT}-solver}. In: Giunchiglia,
  E., Tacchella, A. (eds.) International Conference on Theory and Applications
  of Satisfiability Testing. LNCS, vol. 2919, pp. 502--518. Springer (2003)

\bibitem{minisat-assumptions}
E{\'e}n, N., S{\"o}rensson, N.: {Temporal induction by incremental SAT
  solving}. Electronic Notes in Theoretical Computer Science  89(4),  543--560
  (2003)

\bibitem{FM06}
Fu, Z., Malik, S.: {On Solving the Partial {MAX-SAT} Problem}. In: Biere, A.,
  Gomes, C.P. (eds.) International Conference on Theory and Applications of
  Satisfiability Testing. LNCS, vol. 4121, pp. 252--265. Springer (2006)

\bibitem{graca10}
Gra\c{c}a, A., Lynce, I., Marques-Silva, J., Oliveira, A.L.: {Efficient and
  Accurate Haplotype Inference by Combining Parsimony and Pedigree
  Information}. In: Algebraic and Numeric Biology. pp. 38--56. Springer (2010)

\bibitem{msuncore-aaai11}
Heras, F., Morgado, A., Marques-Silva, J.: Core-guided binary search algorithms
  for maximum satisfiability. In: Burgard, W., Roth, D. (eds.) AAAI Conference
  on Artificial Intelligence. AAAI Press (2011)

\bibitem{hooker-jlp93}
Hooker, J.N.: {Solving the incremental satisfiability problem}. Journal of
  Logic Programming  15(1{\&}2),  177--186 (1993)

\bibitem{jose11}
Jose, M., Majumdar, R.: Cause clue clauses: error localization using maximum
  satisfiability. In: Hall, M.W., Padua, D.A. (eds.) Programming Language
  Design and Implementation. pp. 437--446. ACM (2011)

\bibitem{Kadioglu12}
Kadioglu, S., Malitsky, Y., Sellmann, M.: {Non-Model-Based Search Guidance for
  Set Partitioning Problems}. In: Hoffmann, J., Selman, B. (eds.) AAAI
  Conference on Artificial Intelligence. AAAI Press (2012)

\bibitem{qmaxsat-jsat12}
Koshimura, M., Zhang, T., Fujita, H., Hasegawa, R.: {QMaxSAT: A Partial Max-SAT
  Solver}. Journal on Satisfiability, Boolean Modeling and Computation  8,
  95--100 (2012)

\bibitem{DBLP:conf/sat/2009}
Kullmann, O. (ed.): Theory and Applications of Satisfiability Testing - SAT
  2009, 12th International Conference, SAT 2009, Swansea, UK, June 30 - July 3,
  2009. Proceedings, LNCS, vol. 5584. Springer (2009)

\bibitem{lagerkvist-cp07}
Lagerkvist, M.Z., Schulte, C.: {Advisors for Incremental Propagation}. In:
  Bessiere, C. (ed.) Principles and Practice of Constraint Programming. LNCS,
  vol. 4741, pp. 409--422. Springer (2007)

\bibitem{berre-jsat10}
{Le Berre}, D., Parrain, A.: {The Sat4j library, release 2.2}. Journal on
  Satisfiability, Boolean Modeling and Computation  7(2-3),  59--6 (2010)

\bibitem{manya-handbook09}
Li, C.M., Many{\`a}, F.: {MaxSAT, Hard and Soft Constraints}. In: Handbook of
  Satisfiability, pp. 613--631. IOS Press (2009)

\bibitem{liffiton-jar08}
Liffiton, M.H., Sakallah, K.A.: {Algorithms for Computing Minimal Unsatisfiable
  Subsets of Constraints}. Journal Automated Reasoning  40(1),  1--33 (2008)

\bibitem{Lonsing-corr14}
Lonsing, F., Egly, U.: {Incremental QBF Solving}. Computing Research Repository
  - arXiv  abs/1402.2410 (2014)

\bibitem{mahajan-sat04}
Mahajan, Y.S., Fu, Z., Malik, S.: Zchaff2004: An efficient sat solver. In:
  Hoos, H.H., Mitchell, D.G. (eds.) International Conference on Theory and
  Applications of Satisfiability Testing. LNCS, vol. 3542, pp. 360--375.
  Springer (2004)

\bibitem{manolios-fmcad11}
Manolios, P., Papavasileiou, V.: {Pseudo-Boolean Solving by incremental
  translation to SAT}. In: Bjesse, P., Slobodov{\'a}, A. (eds.) International
  Conference on Formal Methods in Computer-Aided Design. pp. 41--45. FMCAD Inc.
  (2011)

\bibitem{manquinho-sat09}
Manquinho, V., Marques-Silva, J., Planes, J.: {Algorithms for Weighted Boolean
  Optimization}. In: Kullmann  \cite{DBLP:conf/sat/2009}, pp. 495--508

\bibitem{marin-date12}
Marin, P., Miller, C., Lewis, M.D.T., Becker, B.: {Verification of partial
  designs using incremental QBF solving}. In: Rosenstiel, W., Thiele, L. (eds.)
  Design, Automation, and Test in Europe Conference. pp. 623--628. IEEE (2012)

\bibitem{wmsu3-corr07}
Marques-Silva, J., Planes, J.: {On using unsatisfiability for solving Maximum
  Satisfiability}. Tech. rep., Computing Research Repository, abs/0712.0097
  (2007)

\bibitem{martins-ai12}
Martins, R., Manquinho, V., Lynce, I.: {Parallel Search for Maximum
  Satisfiability}. AI Communications  25(2),  75--95 (2012)

\bibitem{martins-sat14}
Martins, R., Manquinho, V., Lynce, I.: {Open-WBO: a Modular MaxSAT Solver}. In:
  International Conference on Theory and Applications of Satisfiability
  Testing. LNCS, Springer (2014)

\bibitem{morgado-constraints13}
Morgado, A., Heras, F., Liffiton, M., Planes, J., Marques-Silva, J.: {Iterative
  and core-guided MaxSAT solving: A survey and assessment}. Constraints  18(4),
   478--534 (2013)

\bibitem{morgado-sat12}
Morgado, A., Heras, F., Marques-Silva, J.: {Improvements to Core-Guided Binary
  Search for MaxSAT}. In: Cimatti and Sebastiani  \cite{DBLP:conf/sat/2012},
  pp. 284--297

\bibitem{nadel-sat12}
Nadel, A., Ryvchin, V.: {Efficient SAT Solving under Assumptions}. In: Cimatti
  and Sebastiani  \cite{DBLP:conf/sat/2012}, pp. 242--255

\bibitem{sinz-cp05}
Sinz, C.: {Towards an Optimal CNF Encoding of Boolean Cardinality Constraints}.
  In: {van Beek}, P. (ed.) Principles and Practice of Constraint Programming.
  LNCS, vol. 3709, pp. 827--831. Springer (2005)

\bibitem{strichman-charme01}
Strichman, O.: {Pruning Techniques for the SAT-Based Bounded Model Checking
  Problem}. In: Margaria, T., Melham, T.F. (eds.) Correct Hardware Design and
  Verification Methods. LNCS, vol. 2144, pp. 58--70. Springer (2001)

\bibitem{vanbeek-hcp06}
{van Beek}, P.: { Backtracking Search Algorithms }. In: Rossi, F., van Beek,
  P., Walsh, T. (eds.) Handbook of Constraint Programming, chap.~4. Elsevier
  (2006)

\bibitem{vanhoeve-hcp06}
{van Hoeve}, W.J., Katriel, I.: Global constraints. In: Rossi, F., van Beek,
  P., Walsh, T. (eds.) Handbook of Constraint Programming, chap.~6. Elsevier
  (2006)

\bibitem{whittemore-dac01}
Whittemore, J., Kim, J., Sakallah, K.A.: {SATIRE: A New Incremental
  Satisfiability Engine}. In: Design Automation Conference. pp. 542--545. ACM
  (2001)

\end{thebibliography}
\bibliographystyle{splncs03}

\end{document}